\documentclass[sn-mathphys-num]{sn-jnl}

\usepackage{graphicx}%
\usepackage{multirow}%
\usepackage{amsmath,amssymb,amsfonts}%
\usepackage{amsthm}%
\usepackage{mathrsfs}%
\usepackage[title]{appendix}%
\usepackage{subcaption}
\usepackage{color}

\usepackage{textcomp}%
\usepackage{manyfoot}%
\usepackage{booktabs}%
\usepackage{algorithm}%
\usepackage{algorithmicx}%
\usepackage{algpseudocode}%
\usepackage{listings}%
\usepackage{natbib}%
\usepackage{multirow}%
\usepackage[table,xcdraw]{xcolor}%
\usepackage{pifont}
\newcommand{\cmark}{\ding{51}}%
\newcommand{\xmark}{\ding{55}}%

\theoremstyle{thmstyleone}%
\theoremstyle{thmstyletwo}%

\theoremstyle{thmstylethree}%

\begin{document}

\title[A Comprehensive Hyperledger Fabric Performance Evaluation based on Resources Capacity Planning]{A Comprehensive Hyperledger Fabric Performance Evaluation based on Resources Capacity Planning}


\author*[1]{\fnm{Carlos} \sur{Melo}}\email{carlos.alexandre@ufpi.edu.br}

\author[1]{\fnm{Glauber} \sur{Gonçalves}}\email{ggoncalves@ufpi.edu.br}

\author[1]{\fnm{Francisco A.} \sur{Silva}}\email{faps@ufpi.edu.br}

\author[1]{\fnm{André} \sur{Soares}}\email{andre.soares@ufpi.edu.br}

\affil[1]{\orgname{Universidade Federal do Piauí}, \orgaddress{\city{Picos}, \state{Piauí}, \country{Brazil}}}


\abstract{Hyperledger Fabric is a platform for permissioned blockchain networks that enables secure and auditable distributed data storage for enterprise applications.
There is a growing interest in applications based on this platform, but its use requires the configuration of different blockchain parameters.
Various configurations impact the system's non-functional qualities, especially performance and cost.
In this article, we propose a Stochastic Petri Net to model the performance of the Hyperledger Fabric platform with different blockchain parameters, computer capacity, and transaction rates.
We also present a set of case studies to demonstrate the feasibility of the proposed model. This model serves as a practical guide to help administrators of permissioned blockchain networks find the best performance for their applications.
The proposed model allowed us to identify the block size that leads to a high mean response time (ranging from 1 to 25 seconds) caused by a change in the arrival rate.}


\maketitle

\section{Introduction}\label{sec1}

Blockchain is a disruptive technology, especially for the productive sector, i.e., industry and services, as it provides resources for secure and decentralized data storage \cite{xu_springer_book2019}. The main objective of blockchain technology is to allow the recording of transactions between two entities, i.e., people or organizations, that may not know each other and thus do not have mutual trust. Among the main properties of blockchain for recording transactions, immutability, audibility, and consistency stand out. These properties result from different unifying technologies for developing distributed systems, such as asymmetric cryptography, consensus protocols, and peer-to-peer networks \cite{greve_minicurso_sbrc2018}.

However, blockchain technology must advance in performance to consolidate its application to the industry, which usually requires low-latency transactions, similar to credit card networks. Such performance must be observed in the widespread public networks Bitcoin and Ethereum \cite{sousa_ijnm2021}.
In this sense, blockchain architectures in private (or permissioned) networks have been proposed \cite{androulaki_eurosys2018}. Hyperledger Fabric is one of the most popular platforms for developing permissioned blockchain networks \footnote{https://www.ibm.com/topics/hyperledger}. It provides capabilities for deploying a blockchain network and running decentralized applications named smart contracts. In this case, network participants form a consortium and share infrastructure costs, seeking better performance than a public blockchain network can provide.

Applications on permissioned blockchains like Hyperledger Fabric have the same security properties as public blockchains, such as consensus protocols, transaction hashing, public-key cryptography, and so on \cite{androulaki_eurosys2018}. 
Additionally, permissioned blockchains offer resources for applications to be accessed exclusively by member users, which extends the use of blockchains to different domains of corporate applications from healthcare \cite{aguiar_fgcs2022} to finance \cite{spunta_2020}. 
However, this type of blockchain is still maturing. 
It needs tools that will allow its adoption, firstly by network administrators and consecutively by software developers for both industry and government.

An essential aspect of permissioned blockchain networks is their performance, particularly the delay of transactions to record data on the blockchain and the throughput of the blockchain network in terms of transactions per second. The Hyperledger Fabric platform, in particular, can achieve different performance levels based on variations in the infrastructure nodes and the blockchain data structure~\cite{guggenberger_cie2022}. The infrastructure nodes are formed by computers organized in a peer-to-peer network that executes procedures in a consensus protocol. Depending on the procedure, they may require different computational capacities. In turn, the blockchain performance is significantly impacted by two parameters: the block size (number of transactions) and the time limit for generating new blocks (\textit{timeout}).

In this work, we propose using Stochastic Petri Net (SPN) to model the performance of the Hyperledger Fabric blockchain with different parameterizations. SPN is known for its high representativeness and is more intuitive than conventional approaches, such as Markov chains~\cite{silva2022model}. Other works have proposed models to analyze availability and costs in deployment~\cite{melo_computing2022,melo_supercomp2021}, identify bottlenecks~\cite{xu_ipm2021,sukhwani2018performance, yuan2020performance}, and the behavior of the network in attack situations~\cite{shahriar2020modelling}. However, these efforts must model the resources available to execute transactions, considering queuing and parallel processing resources. The proposed models do not detail the endorsement, ordering, and \textit{commit} steps.

The proposed SPN model provides valuable resources for configuring and planning permissioned blockchains based on the Hyperledger Fabric platform, which can help increase the industry's adoption of this technology. In this sense, we evaluate the trade-offs between blockchain configurations (block size and block timeout) and the computational capacity of architectural components that impact transaction time and network throughput based on four case studies.

The first case study shows that the Hyperledger Fabric platform requires careful computer capacity adjustment to perform the transaction's final step (\textit{commit}). Otherwise, there may be long response times or unnecessary resource expenses.
In turn, the second case study shows that block size and timeout settings significantly impact the intermediate step of the transaction (ordering). We observed variations in the average response time from 1 to 25 seconds by increasing the block size by just one unit.
Based on these observations, the proposed model allowed us to identify computer node saturation in the initial stages of the transaction (endorsement and ordering), given the interaction between the block size and timeout parameters, which can also explain the increase in latency. Finally, the fourth case study provides a sensitivity analysis that quantifies the impact of both block \textit{timeout} and block size over the system's general performance.

In short, the main contributions of this article are as follows:

\begin{itemize}
\item An analytical model, which is a useful tool for configuring a blockchain on the Hyperledger Fabric platform and understanding the impact of the parameters on the platform performance before its implementation;
\item performance evaluation of the proposed model with parameterized parameters that help us to point out which are the main bottlenecks of the system regarding each step of a blockchain transaction on the Hyperledger Fabric platform;
\item a set of case studies that demonstrates the feasibility of the proposed model by providing a practical guide for performance analysis in blockchain-based networks, as well as a sensitivity analysis evaluation based on the Design of Experiments (DoE) that quantifies which factors impact the most on the system's mean response time (MRT).
\end{itemize}

The remainder of this paper is as follows. Section \ref{sec:related_works} presents some related works and compares them to what we proposed in this paper. Section \ref{sec:background} presents the same basic concepts about the Hyperledger Fabric platform, the system's modeling and Petri nets, and sensitivity analysis through the Design of Experiments. Section \ref{sec:model} provides the proposed performance model that considers each step of a transaction in the Hyperledger Fabric network and which performance metrics we aim to evaluate. Section \ref{sec:res} provides a set of case studies summarizing the results. Finally, Section \ref{sec:conclusão} presents the conclusions and future works.


\section{Related Works}\label{sec:related_works}

In this section, we present some studies that comprises the state-of-the-art in performance models for Hyperledger Fabric, which is the focus of this paper. Recently, several works have aimed to evaluate the performance metrics, availability, and behavior of blockchain systems deployment infrastructures using models, as we discuss now.

Melo et al.~\cite{melo_computing2022,melo_supercomp2021} developed models to analyze the availability and infrastructure resources provisioning for Hyperledger Fabric and Ethereum blockchain platforms. In these works, Continuous Time Markov Chains (CTMC), Reliability Block Diagrams (RBD), and Stochastic Petri Nets were used for modeling those platforms. The authors concluded that the proposed models could help in the planning of blockchain applications based on both platforms, allowing the identification of infrastructure bottlenecks and the comparison of implementation costs. Cost comparisons were also made in implementing infrastructure for them.

Salle et al.~\cite{la2023joint} focus on modeling Hyperledger Fabric (HLF) under cyberattack conditions, specifically through Petri Net simulations of Sybil attacks. In contrast, our paper emphasizes performance optimization across various configurations, examining not only the impact of different network settings on general performance but also exploring how variations in workload affect throughput and latency. While we do not directly address the influence of malicious transactions, our analysis extends to understanding how changes in transaction arrival rates can significantly alter network performance metrics.

The provided abstract focuses on modeling Hyperledger Fabric (HLF) under cyberattack conditions, specifically through Petri Net simulations of Sybil attacks, aiming to fill a gap in academic literature on HLF’s cybersecurity vulnerabilities. This contrasts sharply with the earlier document's broader evaluation of HLF performance through Stochastic Petri Nets without a cybersecurity focus. The new text introduces a novel joint model that quantitatively assesses the impact of cyberattacks on HLF, a specific and targeted analysis that the original study does not undertake. Whereas the first document emphasizes performance optimization across various configurations, the latest abstract highlights innovative contributions to understanding and mitigating cyber risks in blockchain environments, thus marking a significant shift towards enhancing blockchain resilience against cyber threats.

Other works focused on modeling general performance metrics of the Hyperledger Fabric platform specifically~\cite{wu_acm_ease2022, jiang_springer_p2pna2020, ke_springer_cc2022, wang2023stochastic}. 
Jiang et al.~\cite{jiang_springer_p2pna2020} developed a hierarchical model for Hyperledger Fabric 1.4 transactions process and conducted numerical analyses of throughput, discard rate, and mean response time.
Wu et al.~\cite{wu_acm_ease2022} modeled the transaction processes of the Hyperledger Fabric 2.0 using a queuing model with limited transaction pools. By analyzing this model's two-dimensional continuous-time Markov process, they obtained the throughput, transaction rejection probability, system queue length, and mean response time.
Similarly, Ke and Park~\cite{ke_springer_cc2022} developed a series of queuing models for the performance evaluation with different service rates for endorsing and validation through the metrics of queue length and mean response time.

Finally, some performance modeling efforts looked mainly at inferring the metrics throughput and latency of Hyperledger Fabric based on varying arrival rates, block size, and block timeout settings~\cite{xu_ipm2021,sukhwani2018performance, yuan2020performance}. 
Xu et al.~\cite{xu_ipm2021} proposed an analytical model based on equations, and its inferred metrics were compared with simulations. 
Yuan et al.~\cite{yuan2020performance} performed a comparison relying on the Generalized Stochastic Petri Nets (GSPN) modeling approach.
In turn, Sukhwani et al.~\cite{sukhwani2018performance} proposed Stochastic Reward Networks (SRN) models that allowed, in addition to performance metrics, to estimate the average queue size of each step in Hyperledger Fabric transaction flow. Other transaction steps were also analyzed, identifying bottlenecks and presenting several possible scenarios. 
The authors concluded that the time to complete the endorsement step is affected by the number of nodes participating and the policies adopted.

Wang et al.~\cite{wang2023stochastic} delves into the stochastic performance analysis of phase decomposition within Hyperledger Fabric, employing a more granular approach to dissect transaction processing phases—execute, order, validate (endorsing, ordering and committing). It uses stochastical modeling to evaluate how resource allocation, such as CPU and network settings, affects transaction latency and throughput performance metrics. The present paper differs from \cite{wang2023stochastic} by providing a generalizable Petri Net to model the entire transaction flow and interactions within Hyperledger Fabric at each transaction stage in the given system, providing a holistic view of the system's performance under various configurations.

This paper proposes an SPN model to estimate the value of important performance characteristics of Hyperledger Fabric. 
Like some works presented above, the model estimates the metrics of \textit{mean response time, throughput}, and \textit{resource utilization}. 
As additional contributions, the model also estimates two new metrics: the \textit{maximum block size rate} and \textit{timeout block rate}, as well as the \textit{discarding probability}, which directly impacts the efficiency and reliability of transaction processing. 
High discarding probability can lead to increased transaction failures and inefficiencies within the blockchain network, affecting overall system performance and user experience. 
The study also presents a calibration strategy for setting the timeout and block size parameters.

Table~\ref{tab:related} shows the main characteristics addressed by each work discussed in this section. 
We differ from all those prior efforts by jointly considering all these characteristics. 
It is worth mentioning that none of the approaches allowed easily identifying the timeout and size parameters for an arrival rate, as well as to identify block formation failures due to timeout.

\begin{table*}[!htp]
\centering
\begin{center}
\caption{Research Works on Performance Modeling and their comparison to the main aspects addressed by this paper.}
\begin{tabular}{@{}lcccccc@{}}
\toprule
\multicolumn{1}{c}{\textbf{Publication}} & \multicolumn{1}{c}{\textbf{MRT}} & \multicolumn{1}{c}{\textbf{Throughput}} & \multicolumn{1}{c}{\textbf{RU}} & \multicolumn{1}{c}{\textbf{MSBR}} & \multicolumn{1}{c}{\textbf{TBR}} & \multicolumn{1}{c}{\textbf{DP}} \\ \midrule
Sukhwani et al. \cite{sukhwani2018performance} & \cmark & \cmark & \cmark & \cmark & \cmark & \xmark \\
Jiang et al. \cite{jiang_springer_p2pna2020} & \cmark & \cmark & \xmark & \xmark & \xmark & \cmark \\
Yuan et al. \cite{yuan2020performance} & \cmark & \cmark & \xmark & \cmark & \cmark & \xmark \\
Xu et al. \cite{xu_ipm2021} & \cmark & \xmark & \xmark& \cmark  & \xmark & \xmark \\
Melo et al. \cite{melo_supercomp2021} & \xmark & \xmark& \cmark& \xmark  & \xmark & \xmark \\
Ke and Park \cite{ke_springer_cc2022} & \cmark & \xmark& \xmark& \xmark  & \xmark & \xmark \\
Wu et al. \cite{wu_acm_ease2022} & \cmark & \xmark& \xmark& \xmark  & \xmark & \xmark \\
Wang et al. \cite{wang2023stochastic} & \cmark & \cmark & \cmark & \cmark  & \cmark & \xmark \\
Salle et al. \cite{la2023joint} & \cmark & \cmark & \cmark & \xmark  & \xmark & \xmark \\
Melo et al. \cite{melo_computing2022} & \xmark & \xmark& \cmark& \xmark  & \xmark & \xmark \\
This paper & \cmark & \cmark& \cmark& \cmark  & \cmark & \cmark \\ \bottomrule
\end{tabular}
\label{tab:related}
\end{center}
\begin{tablenotes}[flushleft]\footnotesize
\item[]Metrics abbreviations: Mean Response Time (MRT), Resource Utilization (RU), Maximum Size Block Rate (MSBR), Timeout Block Rate (TBR), Discarding probability (DP)
 \par
\end{tablenotes}
\end{table*}

\section{Background}\label{sec:background}

This section presents the main concepts regarding the Hyperledger Fabric platform; system's modeling with Petri nets, and sensitivity analysis through the Design of Experiments.

\subsection{The Hyperledger Fabric Platform}\label{subsec:arq}

The Hyperledger Fabric (HLF) platform is currently one of the most popular blockchain environments. It is an open-source project involving over 35 organizations and 200 developers\footnote{More information can be found at \textit{https://hyperledger-fabric.readthedocs.io}.}.
This blockchain network uses the execute-order-validate strategy to process transaction blocks.
This separation between the execution and ordering processes allows for better scalability and performance compared to order-execute strategies of public blockchains such as Bitcoin and Ethereum~\cite{androulaki_eurosys2018}.

There are many ways to deploy a private or permissioned Hyperledger Fabric infrastructure.
A typical HLF environment has at least two actors: the client and the service provider (network).
In this paper, we evaluate only the service provider side, which means that what happens on the client side does not impact the general performance of the given system.

The network side in an HLF environment usually uses a container-based technology with smart contract (chaincode) management.
Generally speaking, the chain codes manage the business rules and how the application will perform its activities.
These chain codes must be pre-installed in the environment as they are deployed.
Already, the client side requires an SDK written in Javascript, Python, or Java.
The Hyperledger Fabric strategy involves three essential steps for recording transactions on the blockchain: endorsement, ordering, and \textit{commit}.

\begin{figure*}[!htb]
 \center
 \includegraphics[width=0.75\textwidth]{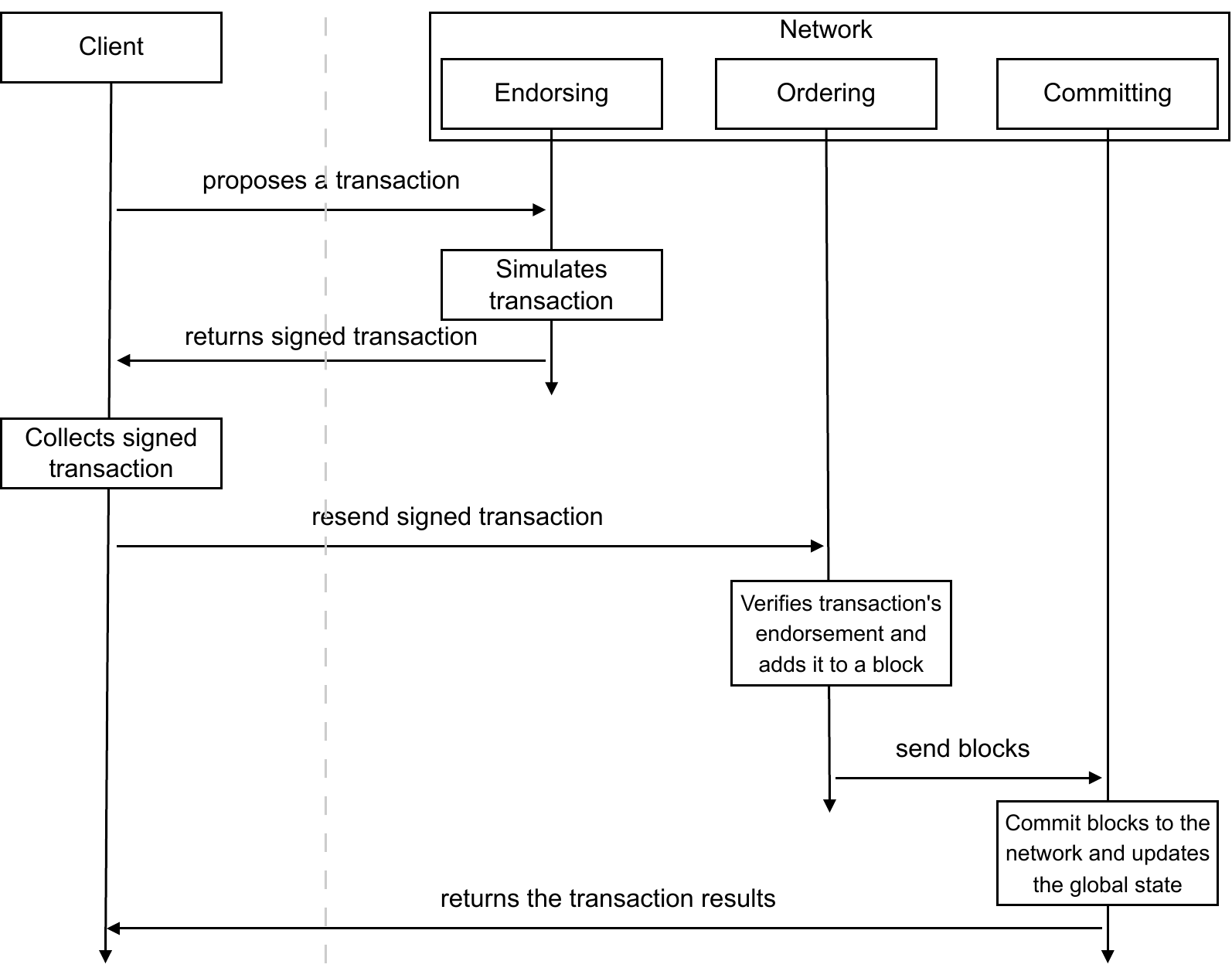}
 \caption{Transaction's flow in the Hyperledger Fabric platform.}
 \label{fig:arch:hlf} 
\end{figure*}

Figure~\ref{fig:arch:hlf} illustrates the transaction flow from the client application to the Hyperledger Fabric network under the three steps mentioned, where each step can be performed on several computers in the network.

Initially, the client application sends the transaction proposal, i.e., a transaction, to a participating computer in the network.
The application waits for endorsements from other participating computers in the network, typically 50\% of the network participants plus one, before sending the transaction to the ordering computer, initiating the \textit{ordering} step.
This step collects network transactions until the defined Batch Timeout (\textit{timeout}) or Batch Size (block size) parameters are reached to generate a new block. Block size parameters can be either by bytes size (PreferredMaxBytes) or by the number of transactions per block (MaxMessageCount); for simplicity, the proposed model only considers the number of transactions per block.
The \textit{endorsement} stage begins, where the transaction execution is simulated on the blockchain, followed by a response to the application with either an endorsement or denial of the transaction.
The new blocks are then sent to the computers responsible for the \textit{commit} step, which consists of updating the global state of the blockchain and persisting transactions onto the ledger.

\begin{figure*}[!htb]
 \center
 \includegraphics[width=0.75\textwidth]{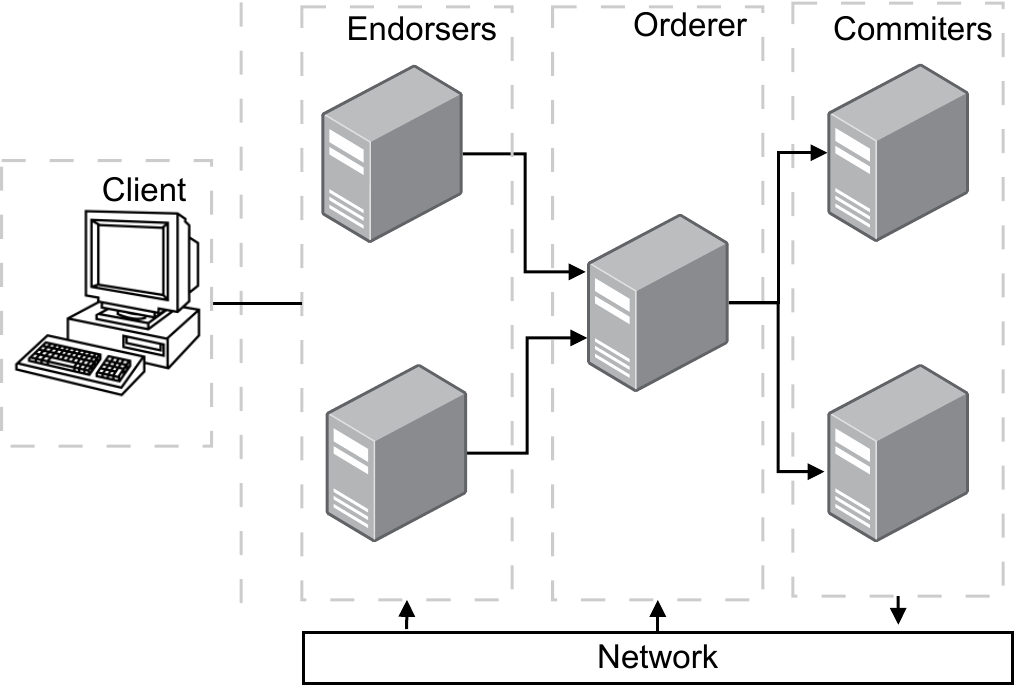}
 \caption{Architecture Overview}
 \label{fig:architecture} 
\end{figure*}

Finally, Figure \ref{fig:architecture} illustrates the modeled architecture, representing the Hyperledger Fabric v2.3, featuring two endorsers, one orderer with Raft consensus mechanism, and two commit nodes. 
This configuration represents a scalable and resilient setup suitable for demonstrating core functionalities while maintaining simplicity for modeling and testing purposes. 

The chosen infrastructure can be generalized to larger deployments, showcasing the advantage of system modeling by allowing comprehensive testing before actual deployment, thereby reducing potential costs and risks associated with scaling up complex blockchain networks.
Also, the endorsers and committers may be the same nodes. 
In contrast, the orderer node is responsible for adding already validated transactions to a block and sending them to persist on the other peers, meaning that a single orderer is enough for most environments.

\subsection{System's Modeling and Petri nets}\label{subsec:petri}

Petri Nets \cite{murata1989} is a family of well-known formalisms for modeling several system types, including concurrency, synchronization, and communication mechanisms, besides supporting deterministic and probabilistic delays.
This paper adopts a particular extension of Petri nets, namely Stochastic Petri Nets \cite{molloy1981}, which allows the association of stochastic delays to timed transitions using different time distributions.

\begin{figure}[!htb]
    \centering
        \includegraphics[width=.81\textwidth]{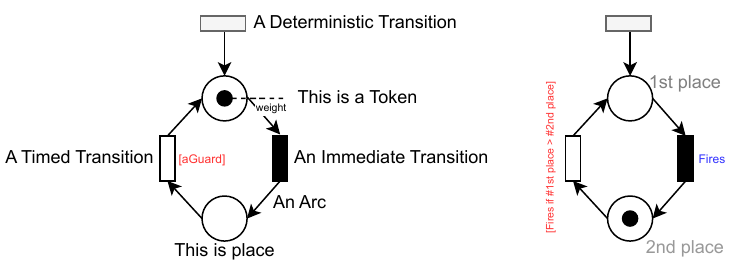}
     \caption{An example of a Stochastic Petri Net}
  \label{fig:example}
\end{figure}

As a high-level definition, a Petri net is a directed bipartite graph composed of places, transitions, tokens, and arcs. 
Arcs connect places to transitions and vice versa, moving tokens from one place to another. 
They may also have weights that indicate the number of tokens required to fire a transition. 
Tokens represent the number of available resources at a given location. Depending on the modeled system and model focus (performance or dependability attribute), they may represent either capacity or quantity. 
Finally, transitions may be named either timed, deterministic, or immediate, and contain guards that associate the firing conditions (moving a token from one place to another) with satisfying some given condition, like a place to have more tokens than another. 
Figure \ref{fig:example} provides an SPN example with the named components and the fire occurrence, which moves a token from 1st to 2nd place.

SPNs are known for their high degree of representativeness and for being more intuitive than conventional options, such as Markov chains, to represent concurrency, parallelism, and synchronization in varied systems~\cite{silva2022model}.

\subsection{Design of Experiments}

Design of Experiments \cite{box1951p} is a method used for conducting Sensitivity Analysis (SA) \cite{matos2015sensitivity, melo2015video}. 
In this method, parameters are referred to as factors, and the assigned value for each factor is known as its level \cite{mathews2005design, matos2015sensitivity}.
This method enables the evaluation of the importance of each system parameter and, additionally, can be used to simultaneously determine the individual and interactive effects of factors that can impact the output measures \cite{mathews2005design, matos2015sensitivity}. 
Examples of analyses conducted to assess the importance of a system's parameters can be found in some previous \cite{franceschini2008model, tiwari2016identification}.

DoE can extract the most information from a small number of experiments, reducing the data collection required \cite{mathews2005design}. Proper analysis of experiments helps to differentiate the effects of various factors that can impact performance \cite{jainart}.
There are several types of DoE, with the most commonly used including full factorial design, fractional factorial design, and simple design \cite{jainart}.

All possible combinations of configurations and workloads are examined in a full factorial design. This method enables the identification of the effect of each factor, including the interactions between them. For a given system that is affected by \texttt{k} factors (parameters), where each factor has \texttt{N} possible levels of values, the number of experiments required would be $N^{k}$ \cite{mathews2005design}. 
As a strategy for dealing with many factors, the number of levels for each can be reduced, such as in a $2^{k}$ factorial design where only two levels are evaluated.

The $2^3$ factorial design involves two levels for each of the three variables, requiring eight runs in total (2 $\times$ 2 $\times$ 2). The design matrix for the $2^3$ design is shown in Table \ref{matrix-doe}. The matrix of runs is generated by alternating between the levels -1 and +1 for the runs of variable $x_3$, followed by alternating pairs of -1s and +1s for variable $x_2$, and finally taking four -1s and four +1s for variable $x_1$. The experiment is balanced, as an equal number of -1s and +1s are in each column. The assignment of the names $x_1$, $x_2$, and $x_3$ to the three columns is arbitrary. The two- and three-factor interactions are included in the table by multiplying the appropriate columns of signs.

\begin{table}[!htb]
\centering
\scriptsize


\begin{tabular}{>{\centering}p{0.5cm}>{\centering}p{0.4cm}>{\centering}p{0.4cm}>{\centering}p{0.4cm}>{\centering}p{0.4cm}>{\centering}p{0.4cm}>{\centering}p{0.7cm}>{\centering}p{0.7cm}>{\centering}p{0.8cm}}
\hline 
\textbf{Std} & \textbf{Run} & \textbf{$x_1$} & \textbf{$x_2$} & \textbf{$x_3$} & \textbf{$x_1$$_2$} & \textbf{$x_1$$_3$} & \textbf{$x_2$$_3$} & \textbf{$x_1$$_2$$_3$}\tabularnewline
\hline 
1 & 8 & - & - & - & + & + & + & -\tabularnewline
2 & 3 & - & - & + & + & - & - & +\tabularnewline
3 & 2 & - & + & - & - & + & - & +\tabularnewline
4 & 6 & - & + & + & - & - & + & -\tabularnewline
5 & 5 & + & - & - & - & - & + & +\tabularnewline
6 & 1 & + & - & + & - & + & - & -\tabularnewline
7 & 4 & + & + & - & + & - & - & -\tabularnewline
8 & 7 & + & + & + & + & + & + & +\tabularnewline
\hline 
\end{tabular}

\caption{Matrix of runs for $2^3$ design.}
\label{matrix-doe}
\end{table}

The experimental runs in Table \ref{matrix-doe} are organized by their logical or standard order indicated by the column labeled \texttt{Std}. 
To prevent the effects of study variables from being confounded with lurking variables, the runs should be performed in random order, such as the order shown in the column labeled \texttt{Run}.

The first run of some random experiment (Run = 1) must be configured with $x_1$ at its +1 level, $x_2$ at its –1 level, and $x_3$ at its +1 level. Then, the combinations of one or more parameters, such as $x_{12}$, $x_{13}$, $x_{23}$, until we have tried all possible combinations together, in this case, $x_{123}$, which will enable us to identify the impact of the interaction between two or more parameters and their given levels.

In this paper, we have applied a $2^{k}$ factorial design to identify the impact of each component in the proposed model on the overall system's performance evaluation, as well as the impact of the interaction between each of these components and their effect on the last step of a transaction in a Hyperledger Fabric network: \textit{commit}.


\section{Proposed Model}\label{sec:model}

This section presents a stochastic model to represent and compute performance characteristics of a Hyperledger Fabric permissioned blockchain network, according to the architecture presented in the previous section.
For this purpose, we specifically explore Stochastic Petri Nets.
The proposed model's objective is to assist systems administrators in the complex task of properly adjusting various parameters to achieve desired performance levels. Therefore, the model should be useful in checking the effect of system changes even before they are implemented.

\begin{figure*}[!htb]
    \centering
        \includegraphics[width=.82\textwidth]{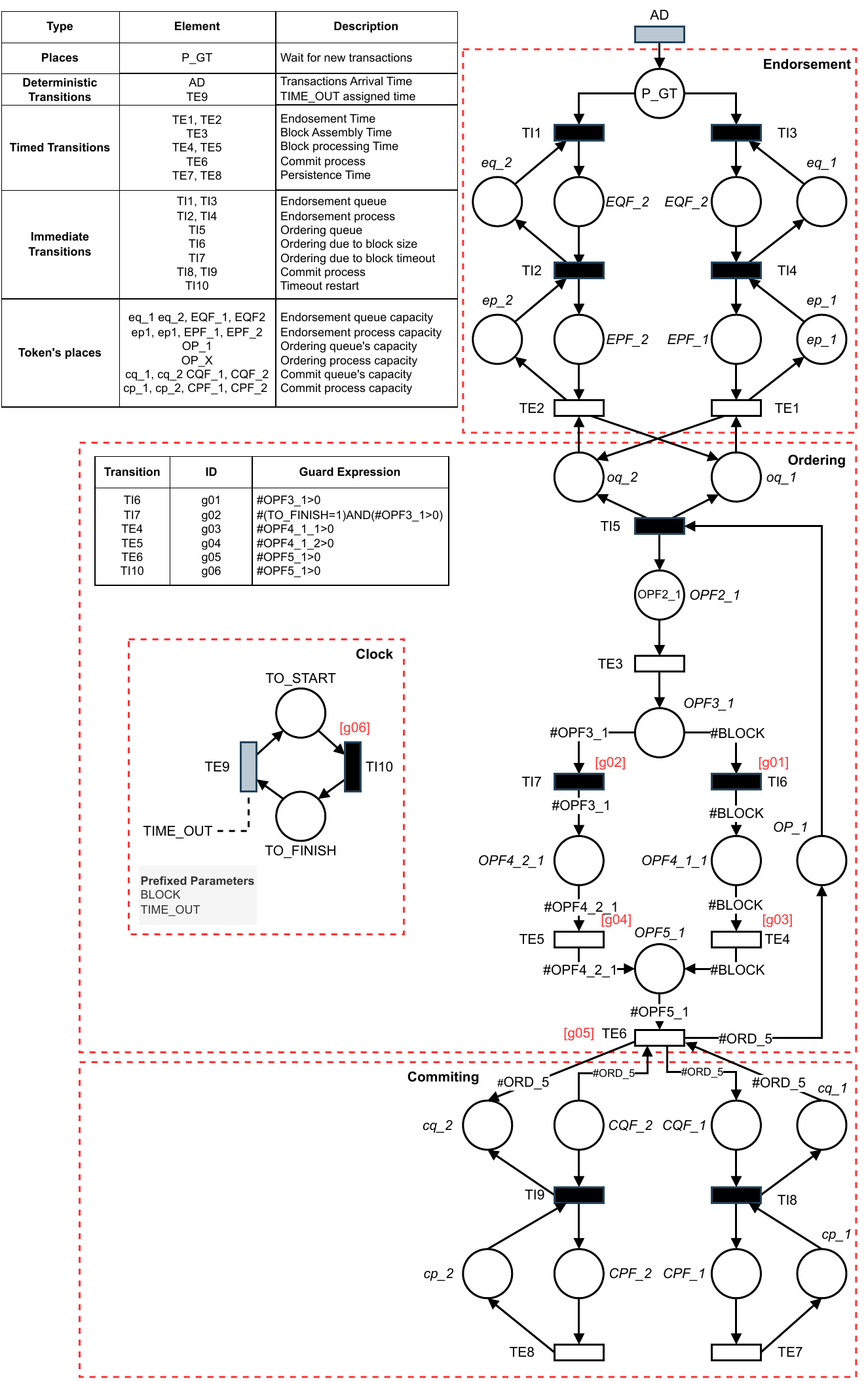}
     \caption{SPN model for processing transactions on Hyperledger Fabric.}
  \label{fig:model}
\end{figure*}

Figure~\ref{fig:model} depicts the proposed SPN model, as well as the description of places, transitions, and the guard expressions for some of these transitions, which are conditions that enable it to fire.
The application generates transactions (i.e., transactions) and sends them to the blockchain network. The network comprises three macro steps, namely \textit{endorsement, ordering}, and \textit{commit}, as shown in the figure from top to bottom.
It is worth noting that these three steps can be performed on $N$ computers.
For brevity, we simplify the description of the SPN model below by assuming that each step is performed on two computers, except for ordering, which is performed on one computer.
A gray transition of deterministic distribution represents the application; a new transaction is generated at each time interval (arrival delay - \texttt{AD}).

The \textit{endorsement} is the first step of the transaction and begins when a token arrives at the location \texttt{P\_GT}.
The OR endorsement policy was considered in the proposed model and respective metrics, meaning that at least one node must sign a given transaction to be considered valid at the ordering step. Changing the metric allows for adopting an AND or a K-out-of-N policy without changing the proposed model. However, for simplicity, the proposed evaluation requires that one of two computers (powered by ~\texttt{T11 or T13}) conduct the endorsement process.
Each computer is represented by two ``triangles" with similar behaviors.
The first triangle illustrates an input queue, and the second is a processing queue.
The input queue does not have an associated time.
Therefore, the transitions \texttt{TI2} and \texttt{TI4} are immediate.
The endorsement times are assigned to the timed transitions \texttt{TE1} and \texttt{TE2}.
The locations at the tops of the triangles represent each computer's queue or processing capacity.
For example, the computer identified with the end $1$ of the endorsement step has some computational nodes (processing cores or containers) equal to \texttt{EP\_1}.
Queuing occurs when the capacity in the triangles is exhausted.
In this example, if the number of processing capacity tokens in \texttt{EP\_1} equals zero, a queue will be at location~\texttt{EQ\_1}.

A single computer is indicated for processing in the next step, \textit{ordering}.
The input queue follows the same pattern as the previous step, with a capacity equal to \texttt{oq\_1}.
The highest risk of bottlenecks occurs at the ordering step, as there are a series of rules described below for the tokens to proceed to the final stage.
Sort processing relies on the ~\texttt{op\_1} capability.
The token effectively enters the sorting step at location~\texttt{OPF2\_1}.
The transition~\texttt{TE3} indicates an individual pre-processing of the transaction to make it able to form a block.
At location~\texttt{OPF3\_1}, tokens accumulate and form a block of two types (complete or partial).
Different block sizes can be configured using the \texttt{\#BLOCK} variable.
If the number of tokens in \texttt{OPF3\_1} is equal to \texttt{\#BLOCK}, these tokens form a complete block and follow their path through the upper transition \texttt{TI6}.
This transition is triggered only if \texttt{\#OPF3\_1} = \texttt{\#BLOCK}.
The model's \texttt{Clock} component serves as a trigger to indicate that the time limit (\textit{timeout}) for forming the complete block has expired.
Thus, the partial block of size \texttt{\#OPF3\_1} goes through the transition \texttt{TI7}, which is triggered only if \texttt{\#TO\_FINISH} = 0.
The blocks (partial or complete) that were formed are finally processed in the transitions \texttt{TE4} and \texttt{TE5}.
When the block arrives at \texttt{OPF5\_1}, the \texttt{Clock} is reset (token changes from \texttt{TO\_FINISH} to \texttt{TO\_START}).

Finally, the \textit{commit} step starts when the formed block reaches the \texttt{TE6} transition and may be performed on all computers allocated for this step through transitions \texttt{TE7} and \texttt{TE8}.

\subsection{Metrics}

The mean response time can be obtained from Little's Law\cite{jainart}, which relates the average number of \text{TransactionsInProgress} in a system, the \text{Arrival\_Rate}, and the \text{MRT}.
As shown earlier, a transaction is subdivided into a set of activities that are directly impacted by the arrival rate.
The arrival rate is the inverse of the arrival time.
Considering the inter-arrival time in the given model, we have $Arrival\_Rate = \frac{1}{\text{Arrival\_Delay}}$. 
It is noteworthy that Little's Law requires a stable system, that is, one that has a transaction rate lower than the server processing rate. 
In the proposed model, we assume that the actual arrival rate is different from the effective arrival rate, as some jobs may be discarded due to the finite size of the queues. 
Then, as recommended in~\cite{jainart}, we subtract the discard rate by the $\text{N\_Discard}$ variable. 
Therefore, the equation corresponding to Little's Law for MRT used in our model is expressed in Equation \ref{eq:little}.

\begin{equation}\label{eq:little}
	\textbf{MRT} = \frac{\texttt{TransactionsInProgress}}{\texttt{Arrival\_Rate}}
\end{equation}

Equation \ref{eq:jobs} defines \texttt{TransactionsInProgress} for the presented model. 
To calculate the number of transactions in progress in the system, we add the number of tokens in each place representing a transaction in progress. 
Here, $E(Place)$ denotes the statistical expectation of having tokens in the given place, where $E(Place) = (\sum_{i=1}^{n} P( m(Place)=i)\times i)$. 
In other words, $E(Place)$ indicates the number of tokens occupying that place.

\begin{equation*}
\label{eq:jobs}
\begin{aligned}
&\textbf{TransactionsInProgress} = \texttt{TransIn\_End+TransIn\_Ord+TransIn\_Com}\\
&\textbf{TransIn\_End} = \texttt{E(P\_GT)+E(EQF\_1)+E(EPF\_1)+E(EQF\_2)+E(EPF\_2)}\\
&\textbf{TransIn\_Ord} = \texttt{E(OQF1\_1)+E($\sum_{i=1}^{n}$OPF)}\\
&\textbf{TransIn\_Com} = \texttt{E(CQF\_1)+E(CPF\_1)+E(CQF\_2)+E(CPF\_2)}\\
\end{aligned}
\end{equation*}

Equation \ref{eq:disc} defines the \texttt{DP\_PROB} (discard rate probability).
It is necessary to have no more queuing capacity left at the system's input.
\texttt{P(Place=n)} calculates the probability that there are $n$ tokens in the given ``Place".
The $\wedge$ symbol indicates a logical AND.

\begin{equation}\label{eq:disc}
	\begin{aligned}
		\textbf{DP\_PROB} = \texttt{P((EQ\_1=0)}\wedge \texttt{(EQ\_2=0))}
	\end{aligned}
\end{equation}

The resource utilization level is calculated for both queue and processing steps.
We focus on using the processing part to exemplify the simulation presented below.
Thus, the utilization is the division of the expected number of tokens in a place (where executed tokens pass) by its total capacity.
For example, the average processing utilization of the endorsement step for the computer with identifier \textbf{1} is given by Equation \ref{eq:u}.
The use of the stage as a whole is given by the average use of computers executed in that stage.

\begin{equation}\label{eq:u}
	\begin{aligned}
		\textbf{U\_END\_1} = \frac{\texttt{E(EP\_1)}}{\texttt{ep\_1}}
	\end{aligned}
\end{equation}

The system flow is represented by the output rate of elements that enter the system.
The exit point of the system is observed, which in our case is the \textit{commit} step, and the exit rate is calculated at that stage. Note that the \textit{commit} step is performed in parallel by $N$ computers. Thus, the outflow is the average of the outflows of the machines participating in the stage.
When we look at the queue or another process step, the throughput is calculated by dividing the expected number of running tokens by the service time of the subsequent timed transition. Therefore, the throughput for computer \textbf{1} can be calculated using Equation \ref{eq:t}, where the time allocated to a transition is given by \texttt{t(Transition)}.

\begin{equation}\label{eq:t}
	\begin{aligned}
		\textbf{TP\_1} = \frac{\texttt{E(CPF\_1)}}{\texttt{t(TE7)}}
	\end{aligned}
\end{equation}

The forking paths from place \texttt{OPF3\_1} provide the transitions trigger rate per timeout or complete block.
These rates follow the same pattern previously exemplified.
Finally, we present two other metrics: BLOCK\_CALL\_RATE and TIME\_OUT\_CALL\_RATE.
Such metrics are respectively given by the equations \ref{eq:c1} and \ref{eq:c2}.
Next, it is possible to analyze the relationship between \textit{timeout} and the block size and calibrate them appropriately for different scenarios.

\begin{equation}\label{eq:c1}
	\begin{aligned}
		\textbf{BLOCK\_CALL\_RATE} = \frac{\texttt{E(OPF\_1\_1)}}{\texttt{t(TE4)}}
	\end{aligned}
\end{equation}

\begin{equation}\label{eq:c2}
	\begin{aligned}
		\textbf{TIME\_OUT\_CALL\_RATE} = \frac{\texttt{E(OPF\_2\_1)}}{\texttt{t(TE5)}}
	\end{aligned}
\end{equation}


\section{Results}\label{sec:res}

In this section, we present four case studies.
Each case study has a specific objective, focusing on one or two parameters.
Case studies are useful for discovering the model's functioning in relation to the real system and illustrating how the model can be explored.
The representation of the model and the computation of the numerical analysis results were obtained with the Mercury~\cite{maciel2017mercury} tool.
The proposed model is available on \href{https://github.com/casm3/hlf-mercury-model}{GitHub} and requires Mercury v5.0.2, that can be found on \href{https://www.modcs.org/?page_id=2392}{MoDCS} Research Group web page
We used the same architecture and number of computers in the case studies presented in Section~\ref{sec:model}.

For model parameters, we initially use the values shown in Table~\ref{tab:cenarioC} for queue capacity (cq), processing capacities for endorsement (ep), ordering (op), and commit (cp) tags and service times in transitions (TE).
Such parameters were based on simulations applied to the proposed model and data provided by previous works, such as \cite{Thakkar2018}, and we considered processing capacities as the number of computers, usually containers on the platform, and queue capacity as the total memory positions for that stage.

\begin{table}[]
\centering
\begin{tabular}{@{}llr@{}}
\toprule
\multicolumn{1}{c}{\textbf{Type}} &
  \multicolumn{1}{c}{\textbf{Parameters}} &
  \multicolumn{1}{c}{\textbf{Values}} \\ \midrule
                           & ep\_1, ep\_2, op\_1, cp\_1, cp\_2 & 6     \\
\multirow{-2}{*}{Capacity} & eq\_1, eq\_2, oq\_1, cq\_1, cq\_2 & 100   \\ \midrule
                           & TE1, TE2, TE3                     & 0.005 s  \\
                           & TE6                               & 0.01 s \\
                           & TE7, TE8                          & 0.08 s \\
\multirow{-4}{*}{Time}     & TE4, TE5                          & 0.002 s  \\ \bottomrule
\end{tabular}
\caption{Initial model configuration parameters used in case studies.}
\label{tab:cenarioC}
\end{table}

The processing capacity values (ep\_1, ep\_2, op\_1, cp\_1, cp\_2) represent assumptions about the resources available for the endorsement, ordering, and commit processes. 
Similarly, the queue capacity (cq\_1, cq\_2, eq\_1, eq\_2, and oq\_1) was estimated as a theoretical allocation of memory positions for each stage of the transaction process.


\subsection{Case Study 01 - Committer Capacity}

In the first case study, we varied the processing capacity of the computers in the \textit{commit} stage.
We tested three possibilities for the number of containers available on each computer.
The tags in the template responsible for processing in \textit{commit} are respectively \texttt{cp\_1} and \texttt{cp\_2}.
The values for such tags were varied as follows: \texttt{cp\_1} = \texttt{cp\_2} = $[2, 4, 6]$.
To obtain a macro view, we also varied the arrival rate with a minimum value of 2.5 transactions per second (tps) a maximum value of 200 tps, and an increment of 15 tps.

For analysis purposes, we use a transaction per block, so we set a high \textit{timeout} (\texttt{TIME\_OUT}=10000ms) and a low block size (\texttt{BLOCK}=1).
With such values, the entire model flow goes through full block formation.

\begin{figure}[htpb]
\centering
\subfloat[]{ \includegraphics[width = 1.6in]{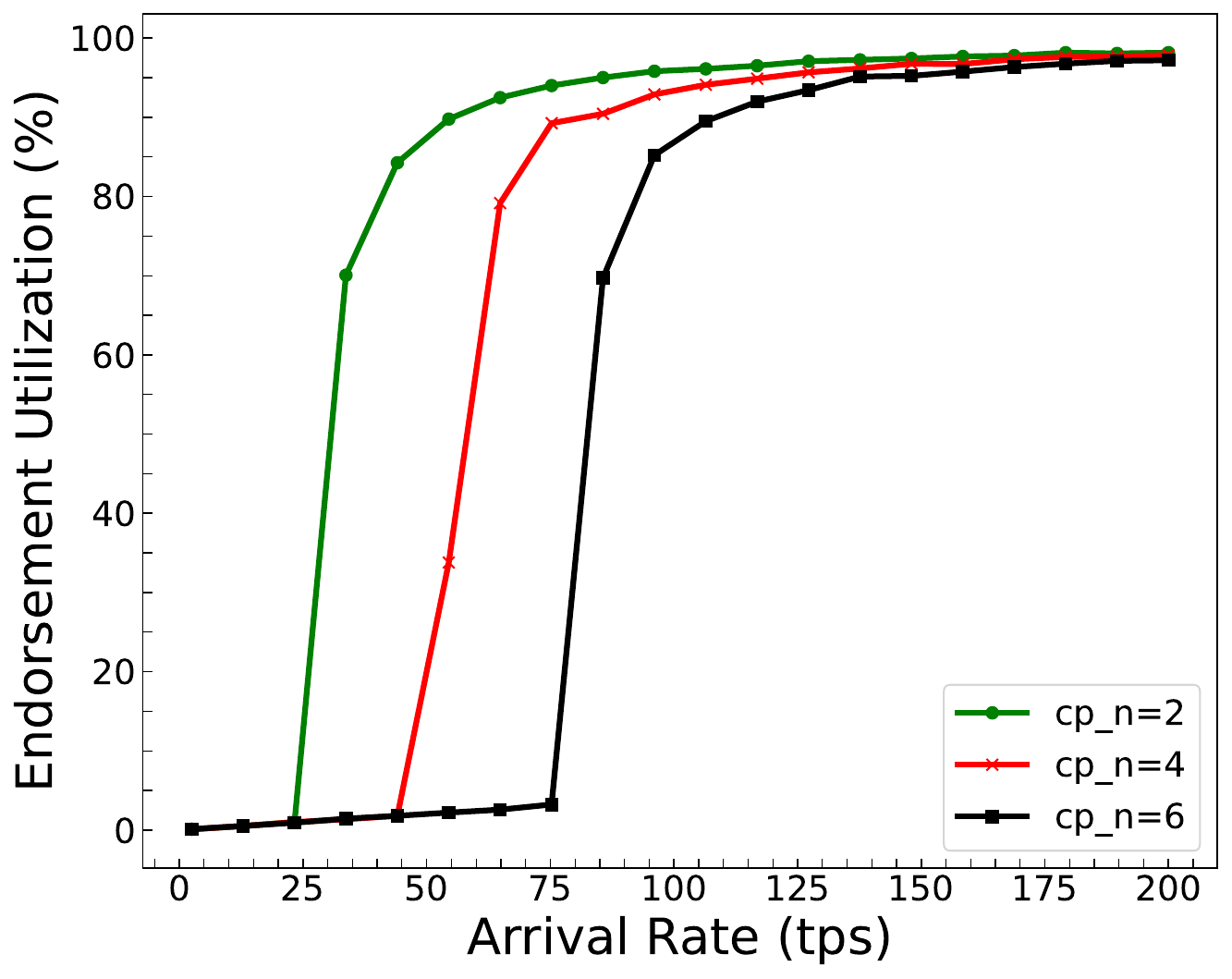}\label{fig:AR_U_end}}
\subfloat[]{ \includegraphics[width = 1.6in]{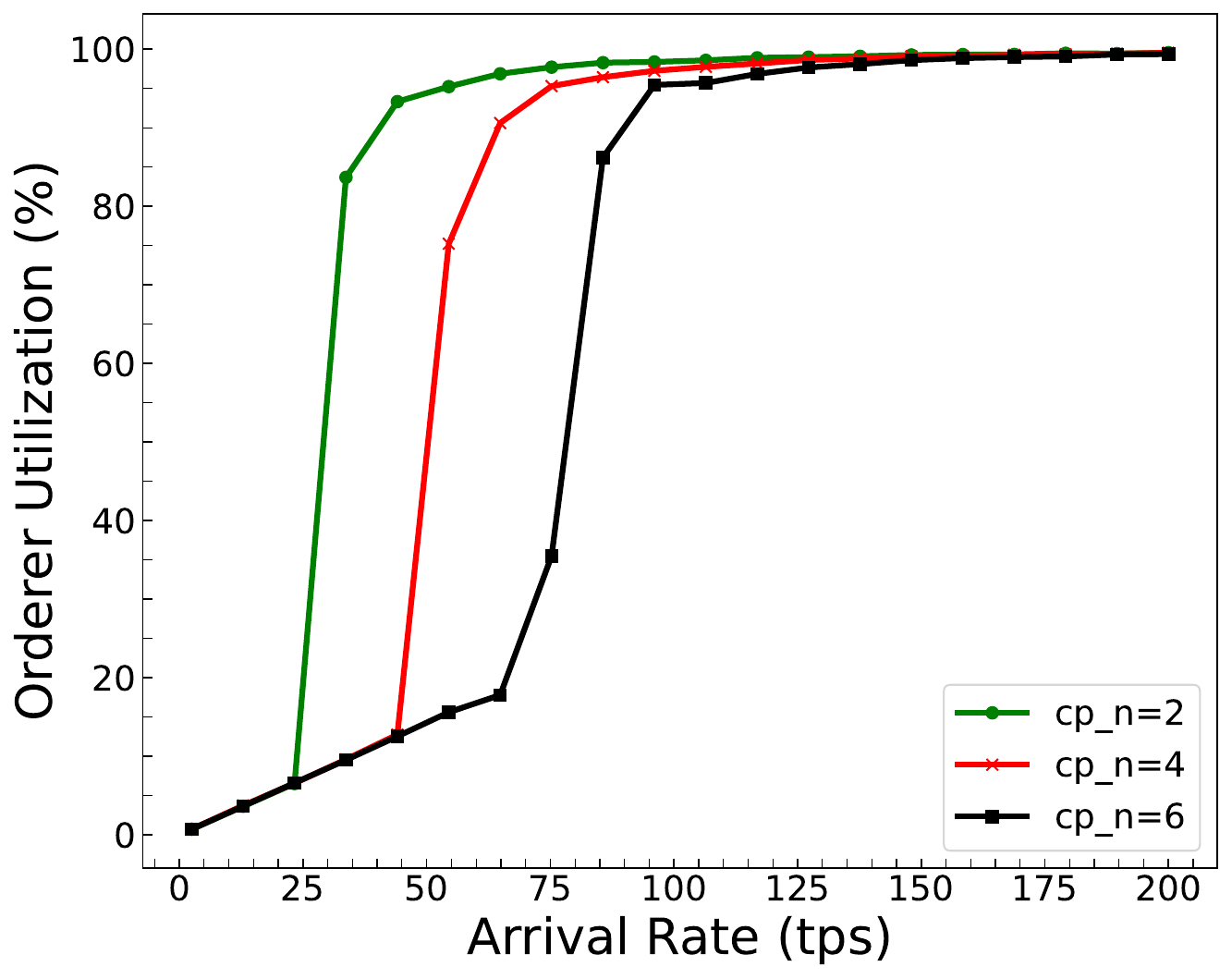}\label{fig:AR_U_Ord}}
\subfloat[]{ \includegraphics[width = 1.6in]{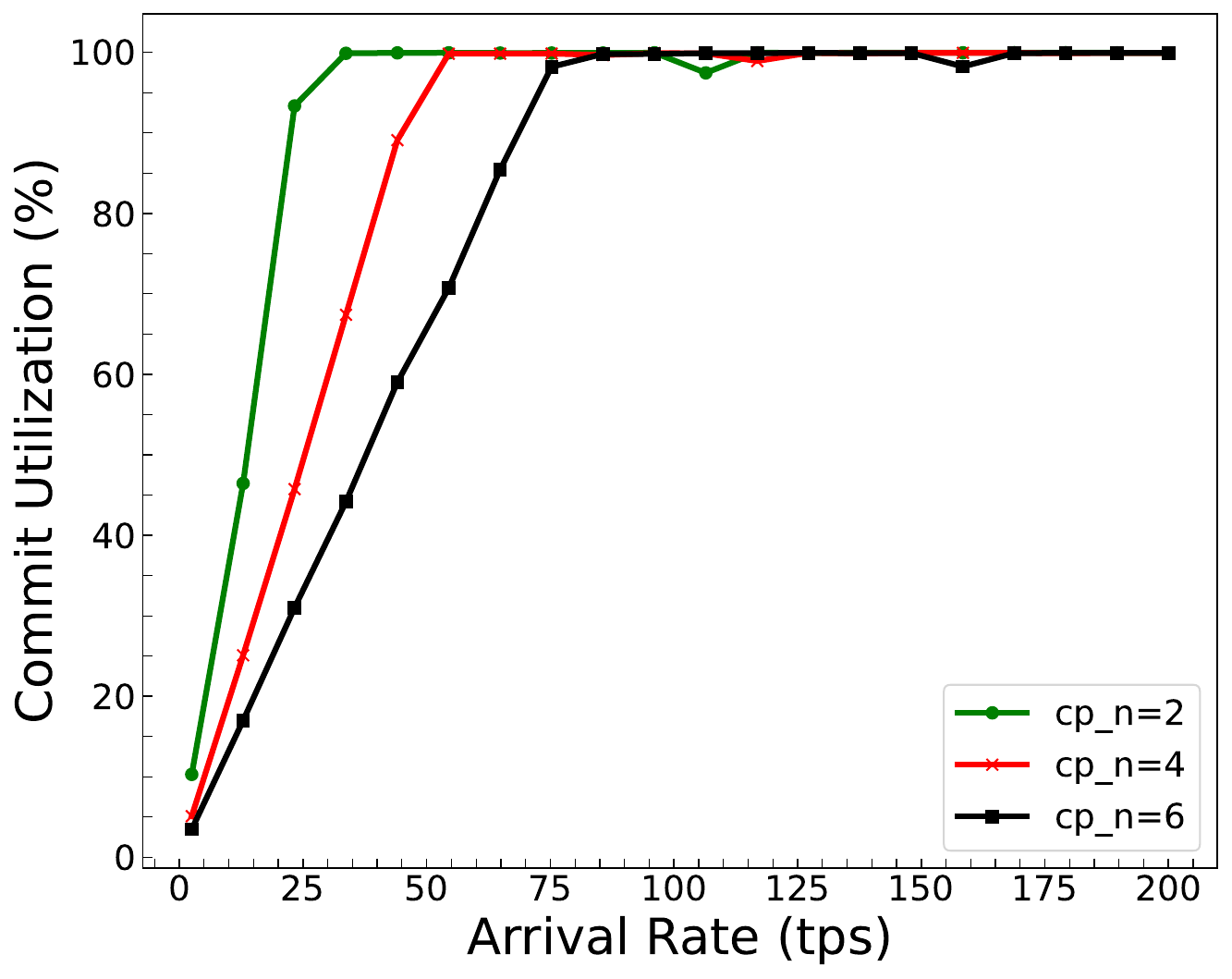}\label{fig:AR_U_Com}} 

\subfloat[]{ \includegraphics[width = 1.6in]{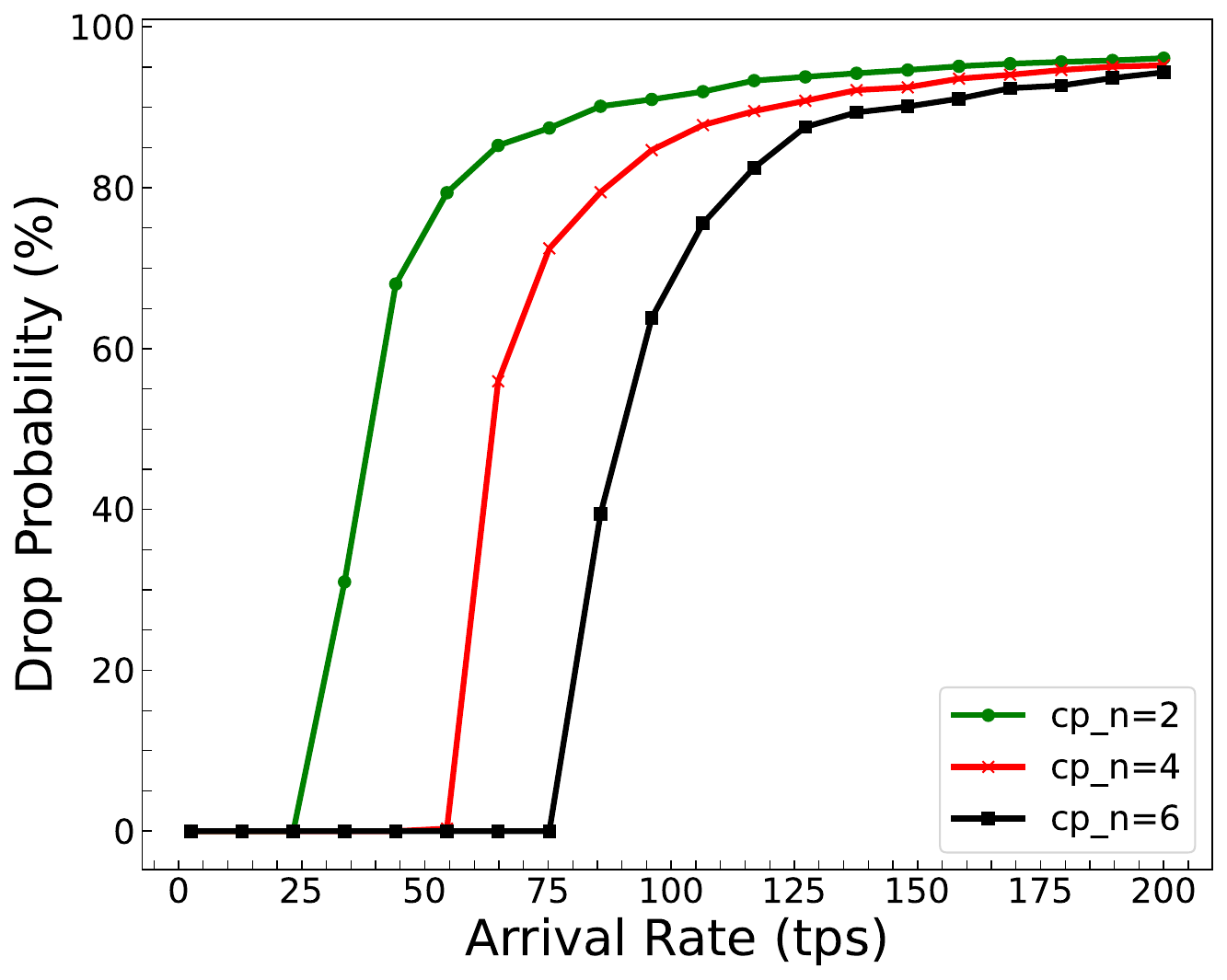}\label{fig:AR_DP}}
\subfloat[]{ \includegraphics[width = 1.6in]{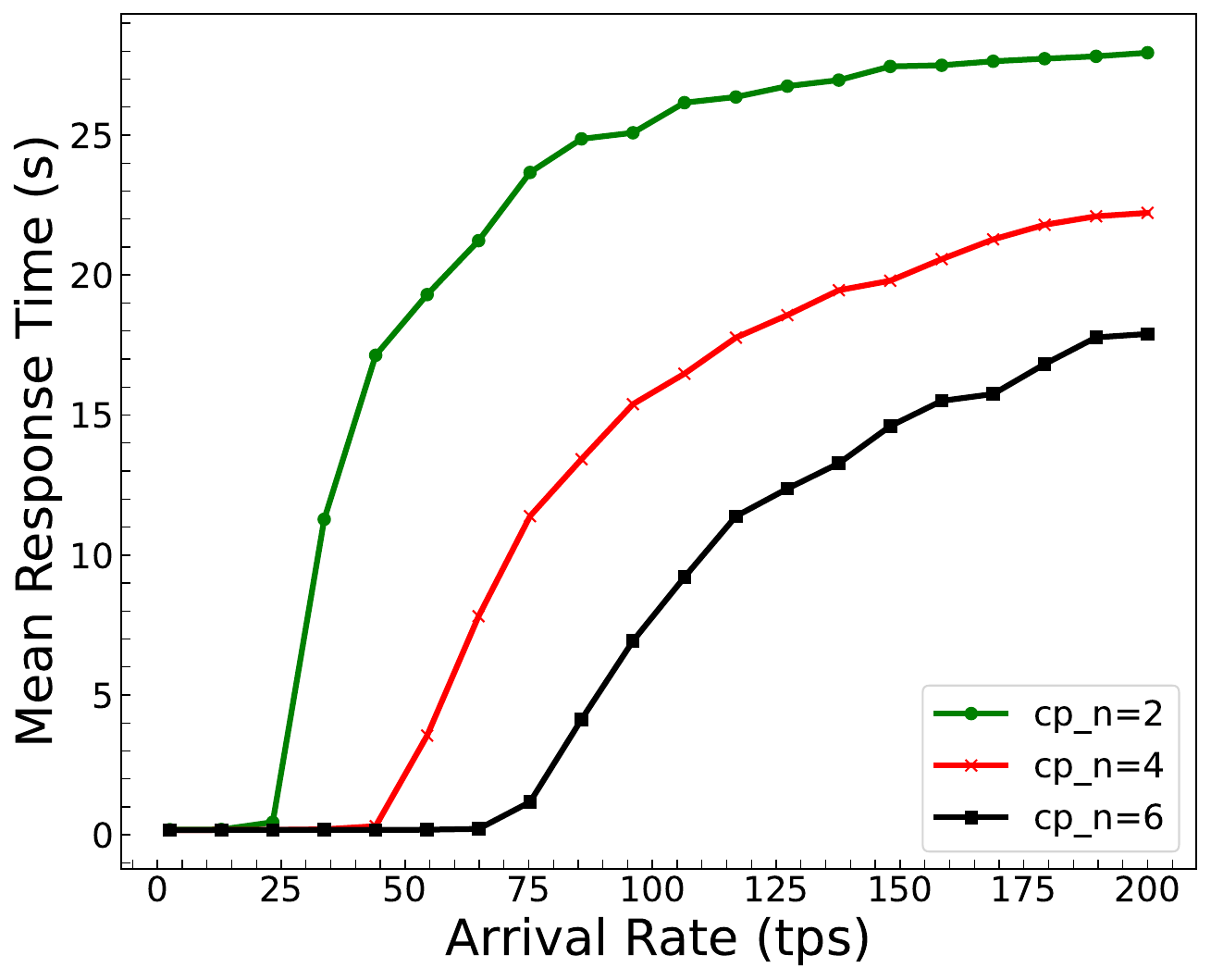}\label{fig:AR_MRT}} 
\subfloat[]{ \includegraphics[width = 1.6in]{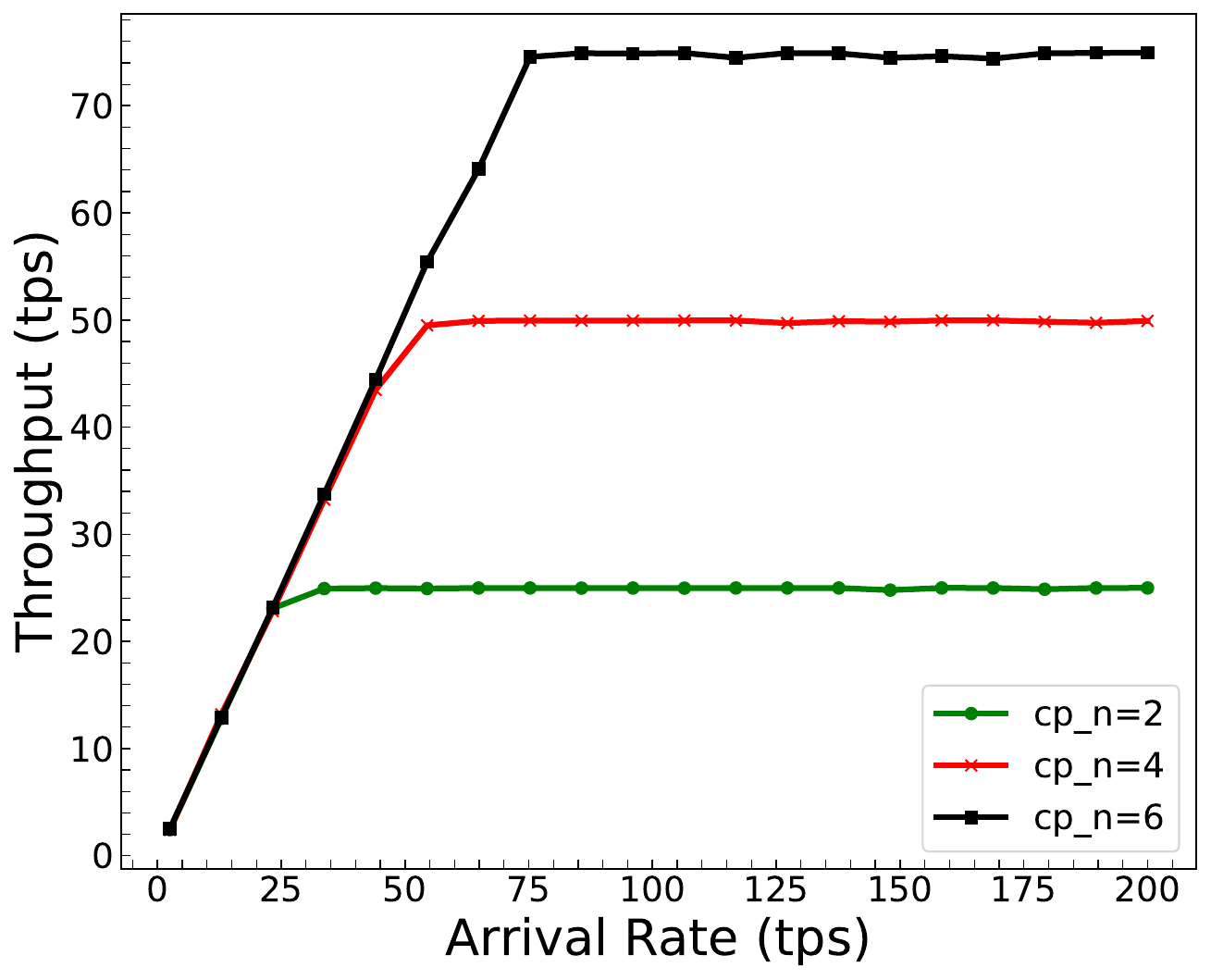}\label{fig:AR_TP}}

\subfloat[]{ \includegraphics[width = 1.6in]{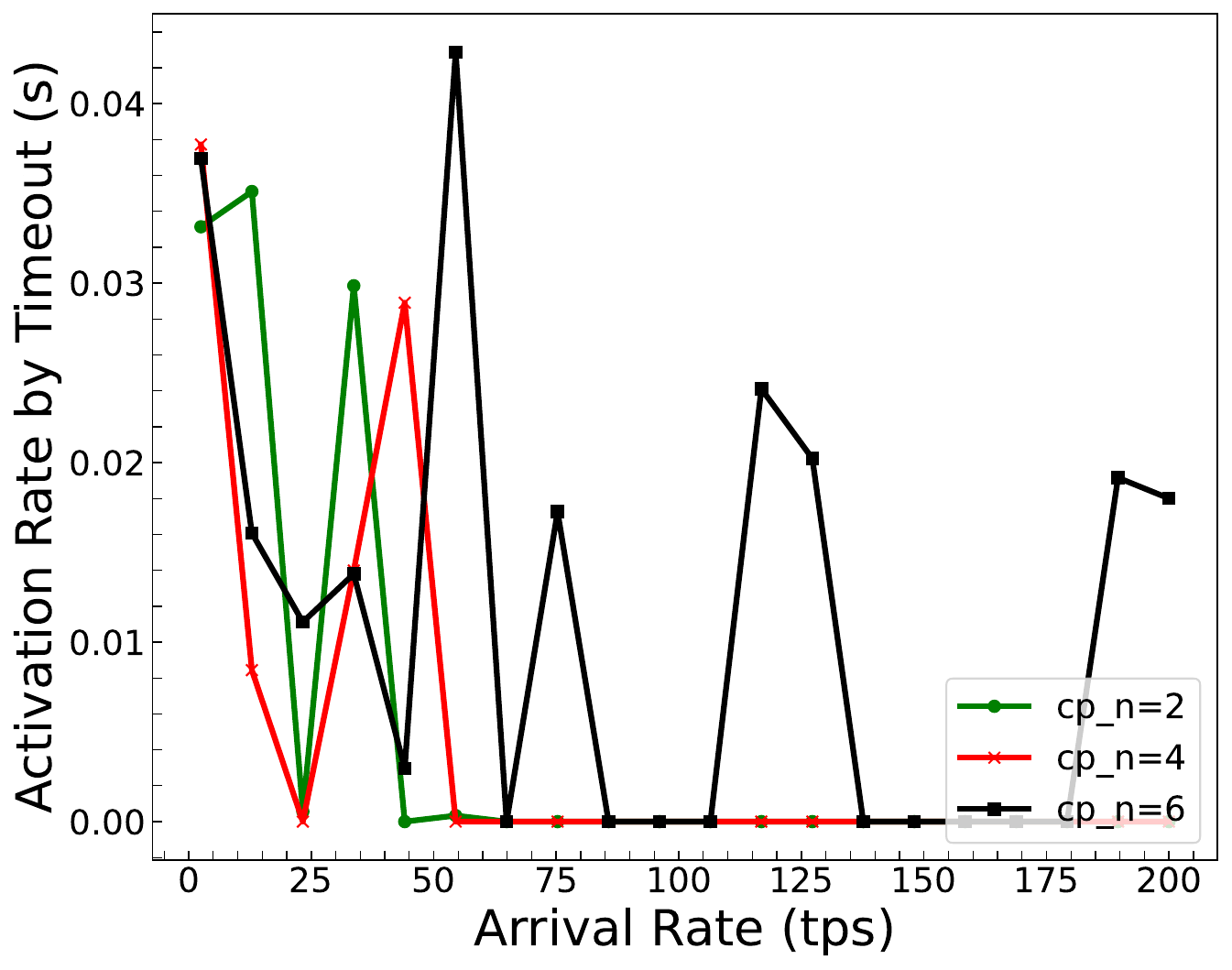}\label{fig:AR_TCALL}}
\subfloat[]{ \includegraphics[width = 1.6in]{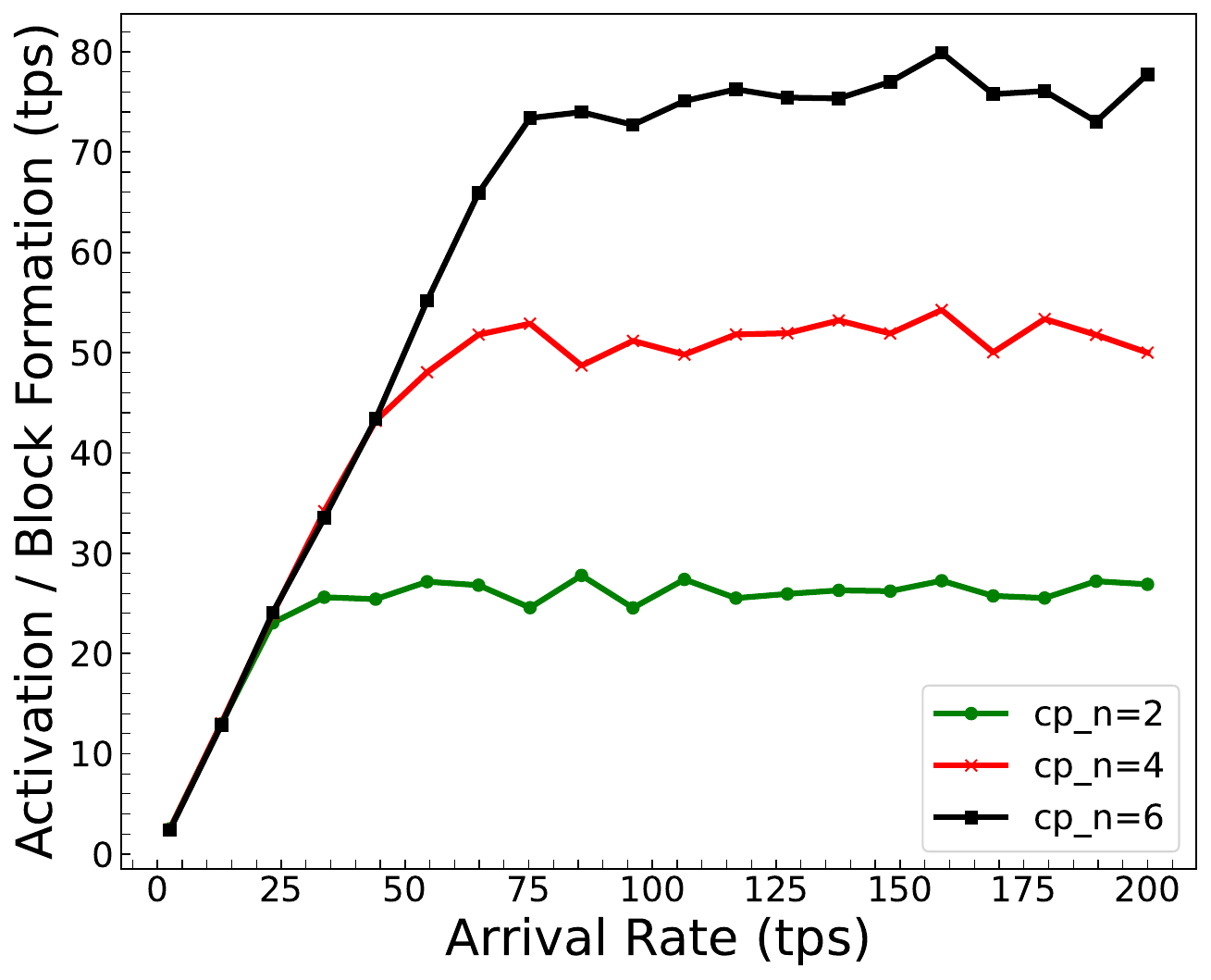}\label{fig:AR_BCALL}} 

\caption{Case Study 01 - Committer Capacity -
(a) Usage - Endorsement
(b) Usage - Ordering
(c) Usage - Commit
(d) Discard Rate 
(e) MRT
(f) Flow
(g) Timeout Activation
(h) Size Activation Block
}
\label{fig:sc1}
\end{figure}

Figure \ref{fig:sc1} presents the results of case study 01, where eight graphs show the impact of the variation of each parameter.
The three utilization graphs (Figure \ref{fig:sc1}a, \ref{fig:sc1}b and \ref{fig:sc1}c) have similar growth behaviors as a function of the arrival rate increase.
However, the usage of \textit{commit} grows quickly due to changing compute capacity.
Another relevant observation is that utilization reaches it peaks for 75 tps in the \textit{commit} step.
This observation is useful for equipment budgets more suited to a specific demand.

The discard rate (Figure \ref{fig:sc1}d) is calculated based on the system entry point, the endorsement step.
Thus, it is observed that Figure \ref{fig:sc1}a and Figure \ref{fig:sc1}d are similar quite similar.
If the endorsement component reaches a high level of utilization, all internal components of the system will also be overloaded, and thus there will be a high probability of discards.

For the MRT of \texttt{cp\_n} = 2, the general performance decreases as the arrival rate increases. As for the MRT (Figure \ref{fig:sc1}e), comparing \texttt{cp\_n} = 4 and \texttt{cp\_n} = 6, for an arrival rate equal to 50 tps, the difference reaches nearly five seconds.

The flow (Figure \ref{fig:sc1}f) grows until it stagnates.
Stagnation occurs due to the use of the commit (Figure \ref{fig:sc1}c).
Once the commit utilization reaches 100\% utilization, there is no way to increase it further.
The activation rate per timeout (Figure \ref{fig:sc1}g) has values well below zero because we configured a timeout of 10s so that the data flow would not follow this path but the path of the complete block formation.
The activation rate per block (Figure \ref{fig:sc1}h) would be proportional to the arrival rate if there were no resource constraints, especially in the \textit{commit} step.
The trigger rate per block with \texttt{cp\_n} = 2 is lower than the others because fewer transactions pass given this configuration.

As the main findings of this case study, we can highlight that the \textit{commit} step plays a fundamental role and must have its capacity carefully configured.
If there is a high restriction on its capacity, the system's performance will drop quickly, reflecting on all metrics.
However, despite having a proportional increase in the three analyzed capacities, we have that the MRT with cp\_n=4 and cp\_n=6 are very similar, and therefore, for the selected arrival rate, the analyst can choose to use a cp\_n=4 capability and save resources.

\subsection{Case Study 02 - Individual Impacts of Block Size and Time Out}

\begin{figure*}[htpb]
\centering
\subfloat[]{ \includegraphics[width = 1.6in]{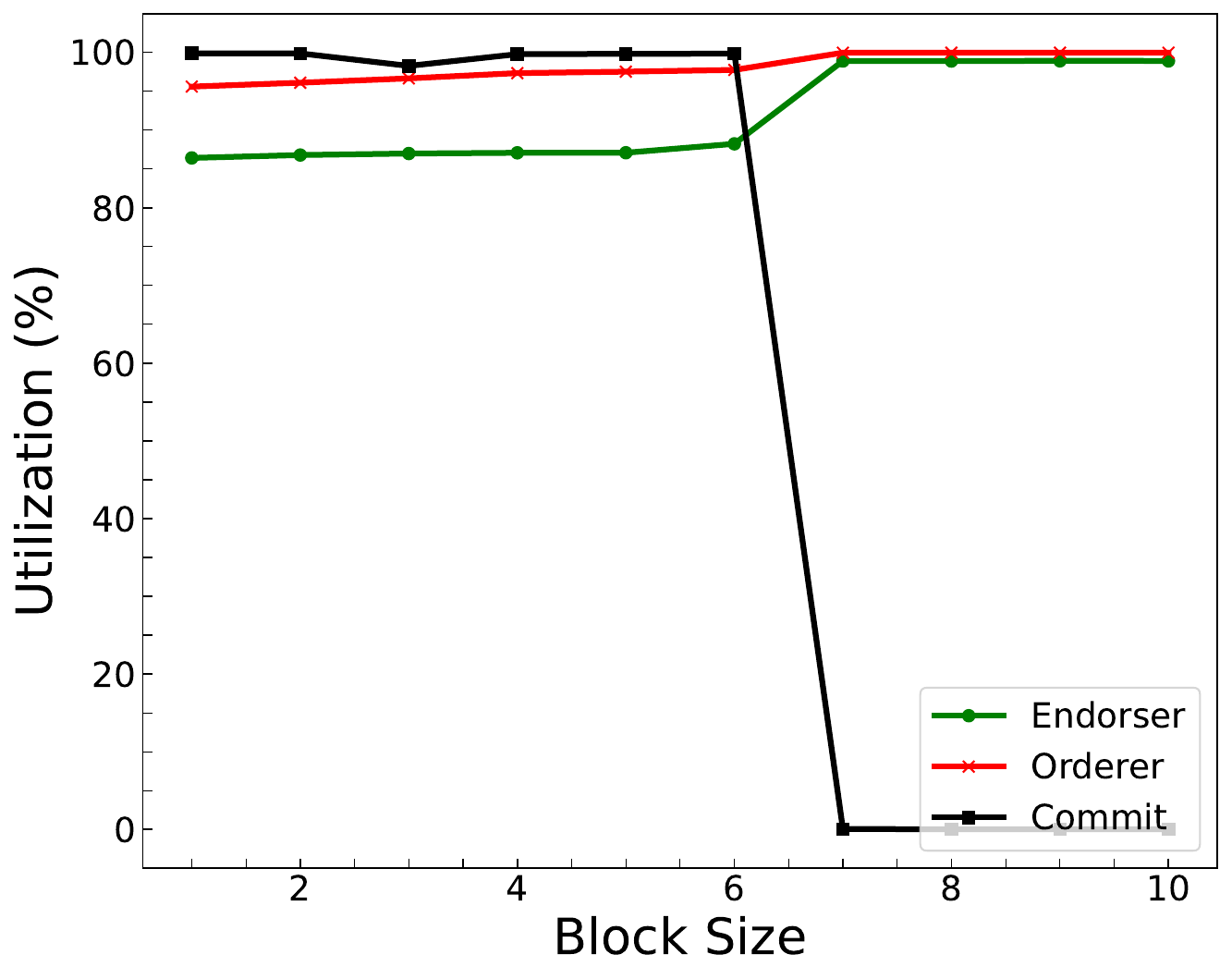}\label{fig:Blo_U}}
\subfloat[]{ \includegraphics[width = 1.6in]{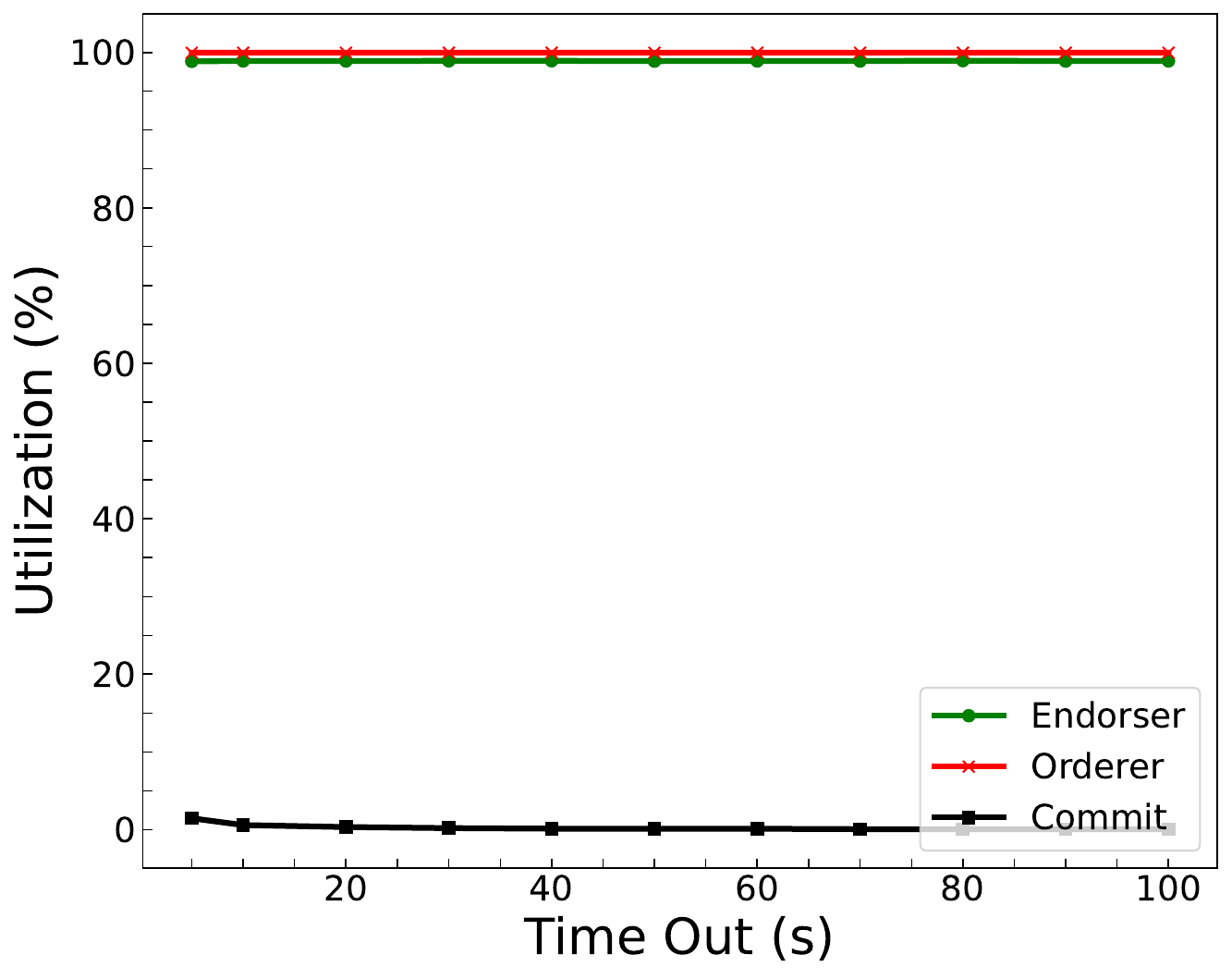}\label{fig:Tim_U}} 
\subfloat[]{ \includegraphics[width = 1.6in]{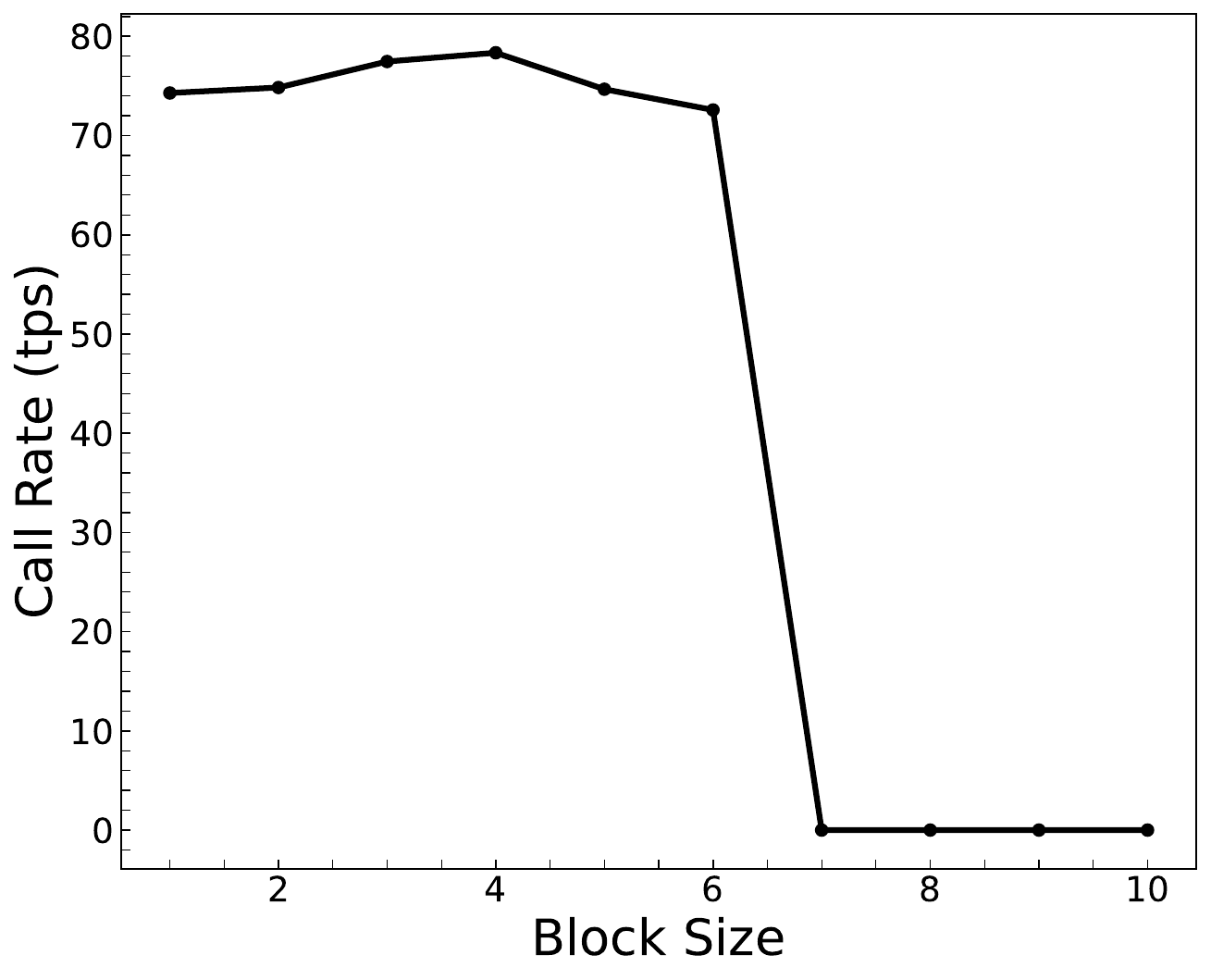}\label{fig:Blo_Call}}

\subfloat[]{ \includegraphics[width = 1.6in]{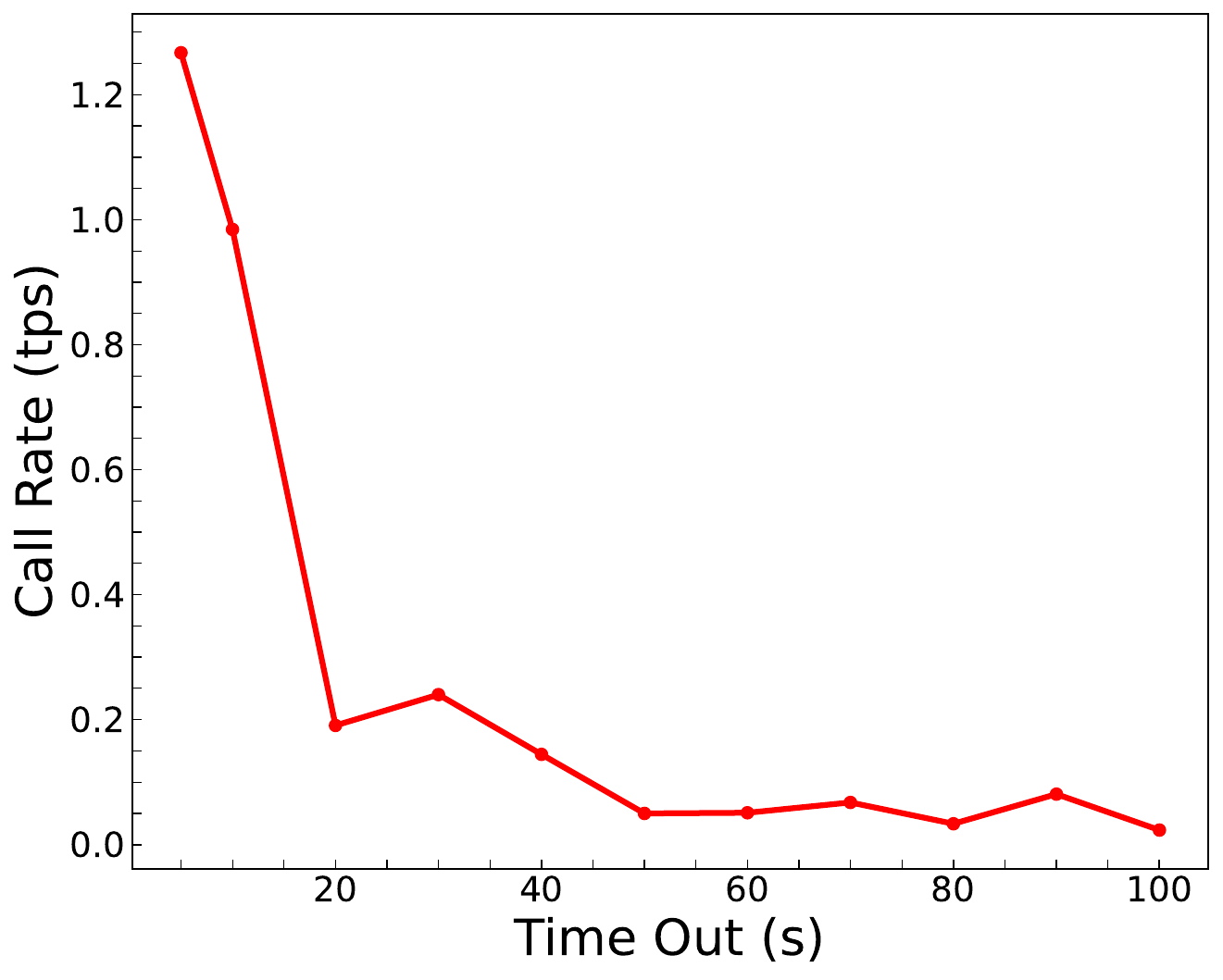}\label{figTim_Call}}
\subfloat[]{ \includegraphics[width = 1.6in]{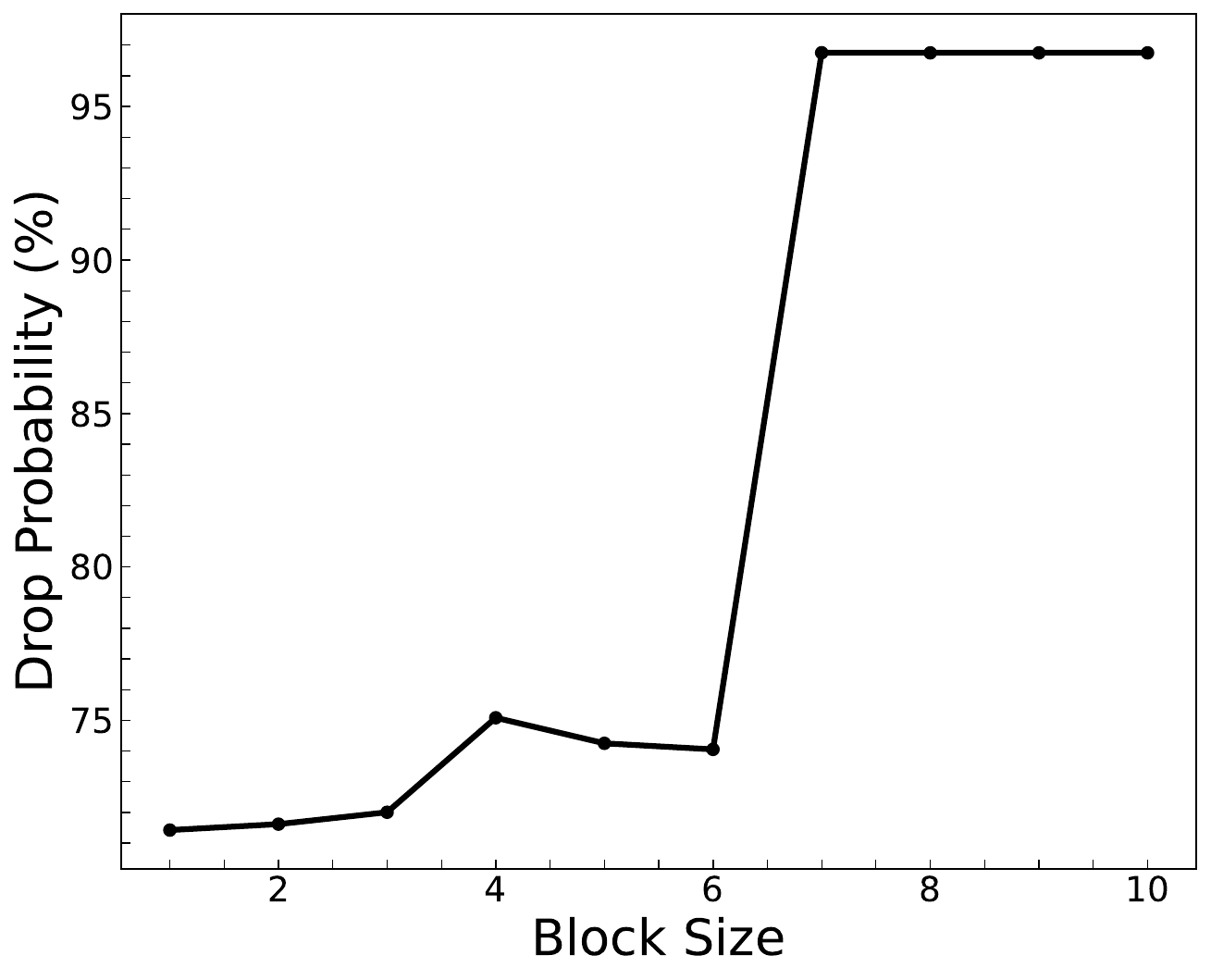}\label{fig:Blo_DP}} 
\subfloat[]{ \includegraphics[width = 1.6in]{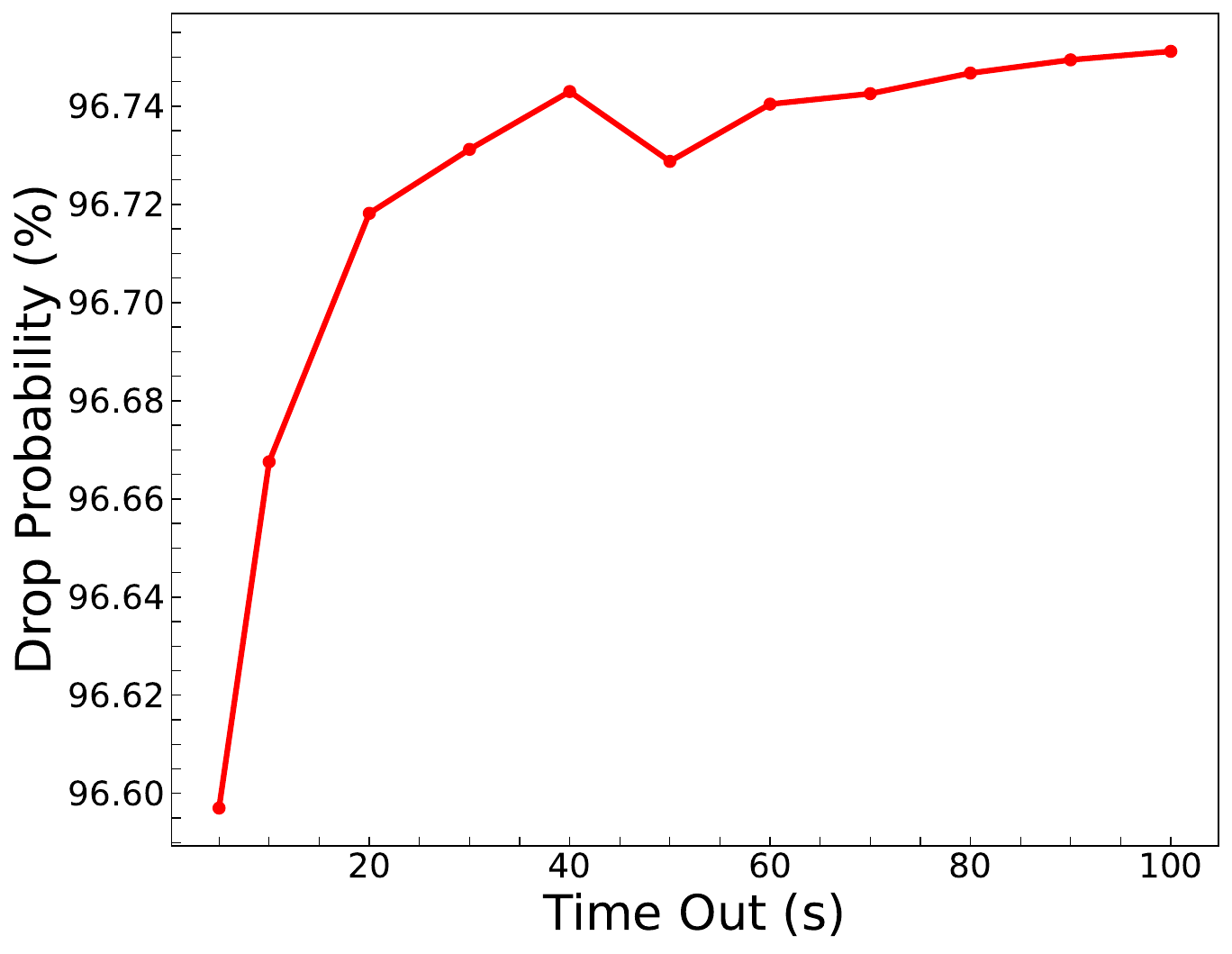}\label{fig:Tim_DP}}

\subfloat[]{ \includegraphics[width = 1.6in]{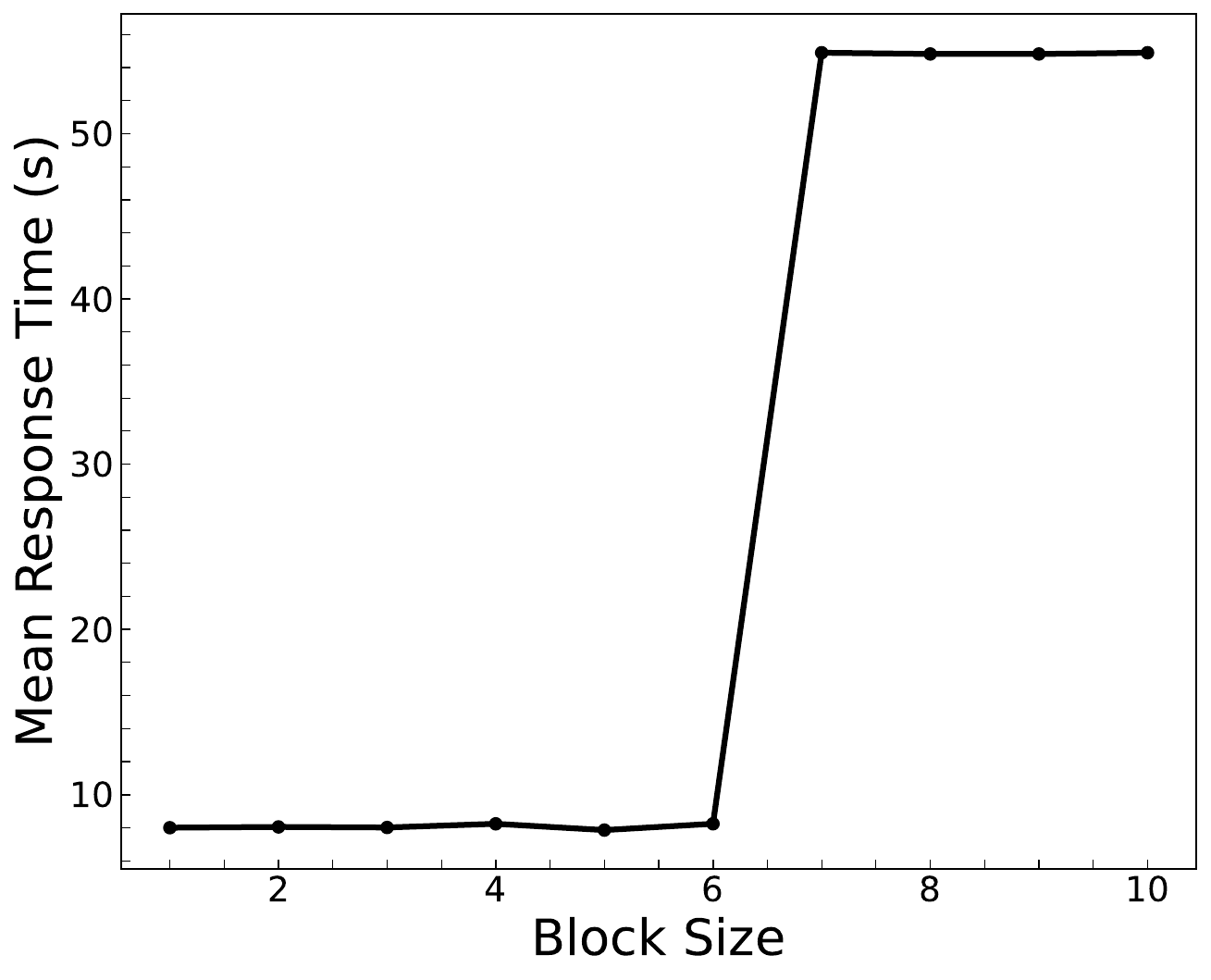}\label{fig:Blo_MRT}} 
\subfloat[]{ \includegraphics[width = 1.6in]{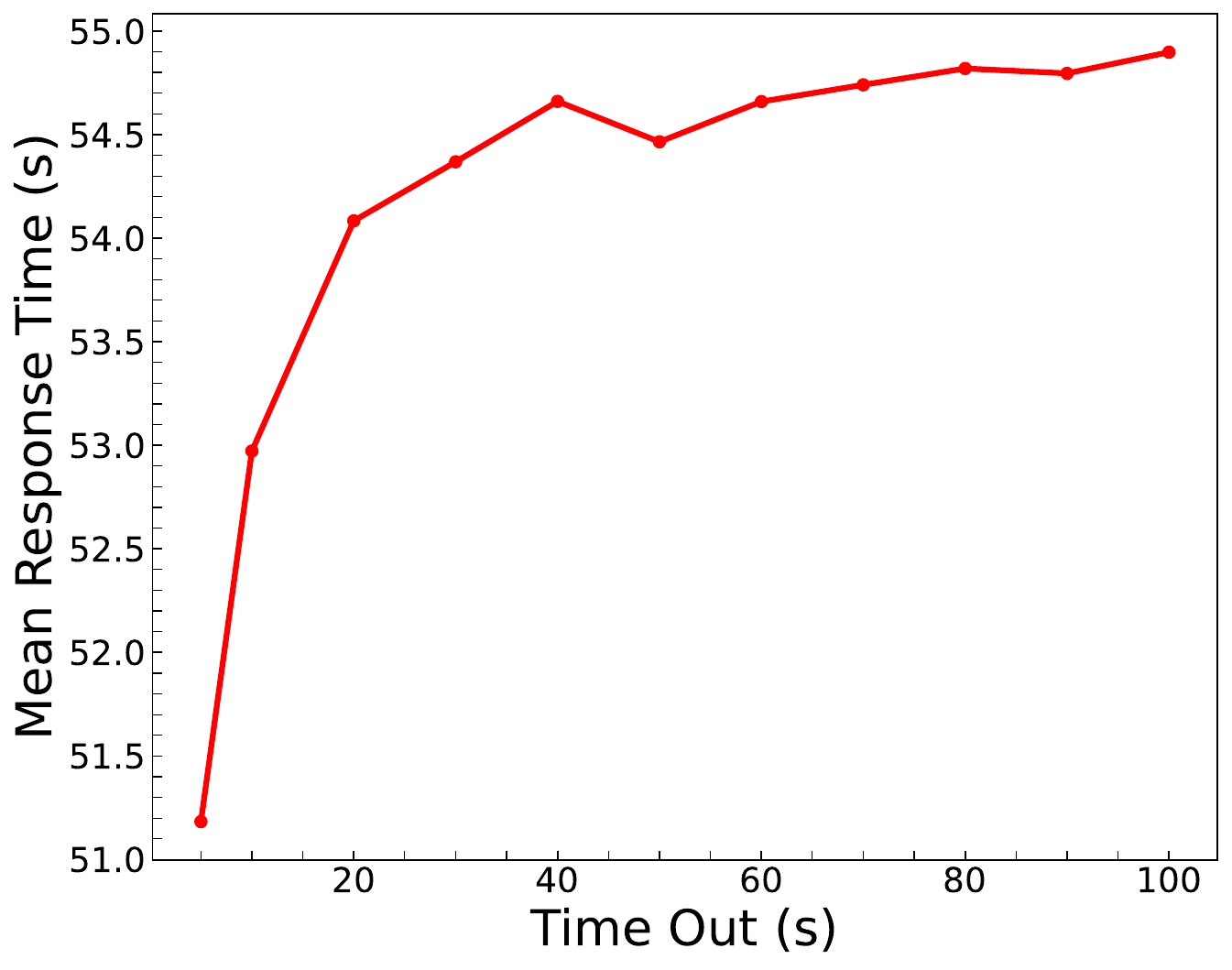}\label{fig:Tim_MRT}} 
\subfloat[]{ \includegraphics[width = 1.6in]{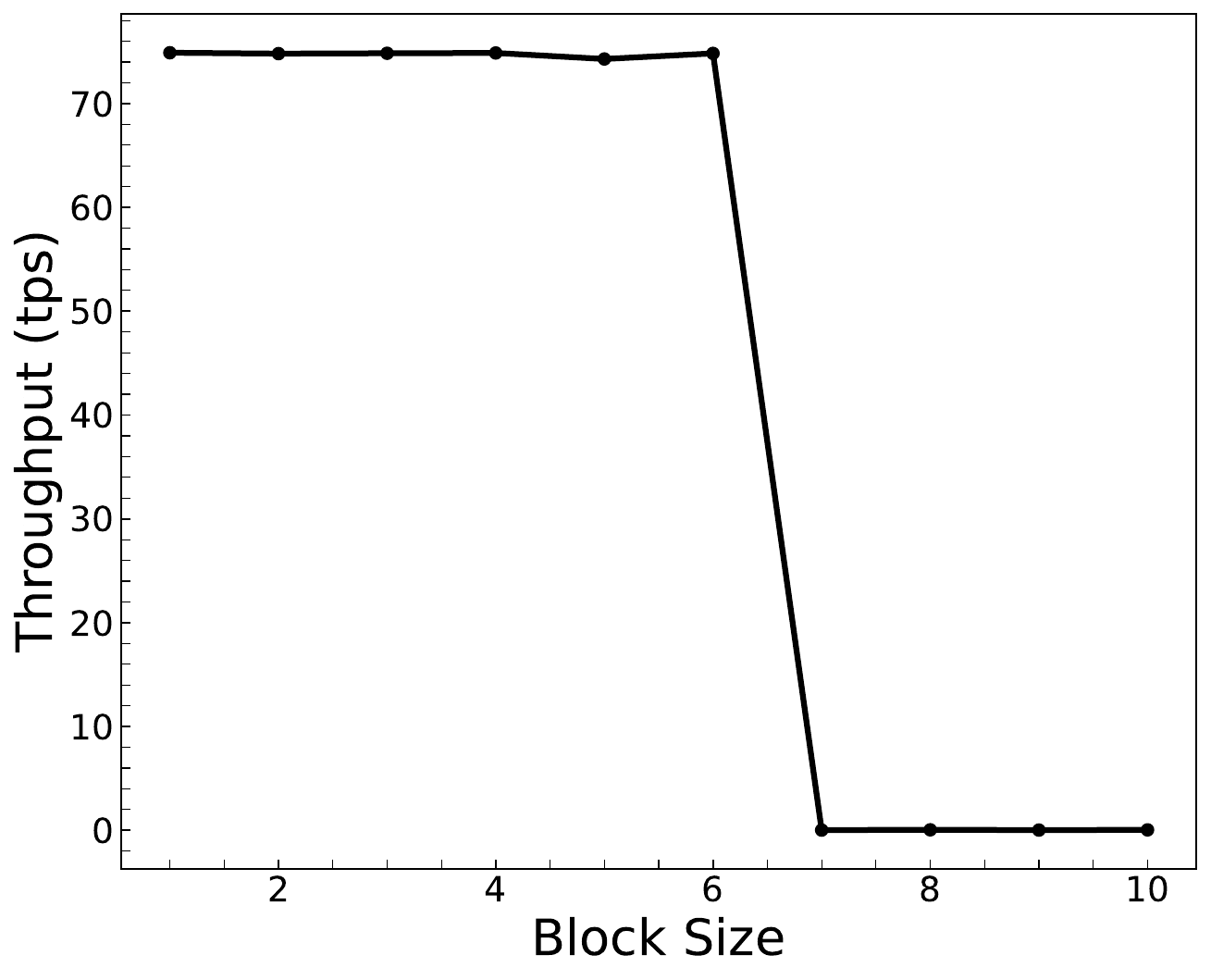}\label{fig:Blo_TP}} 

\subfloat[]{ \includegraphics[width = 1.6in]{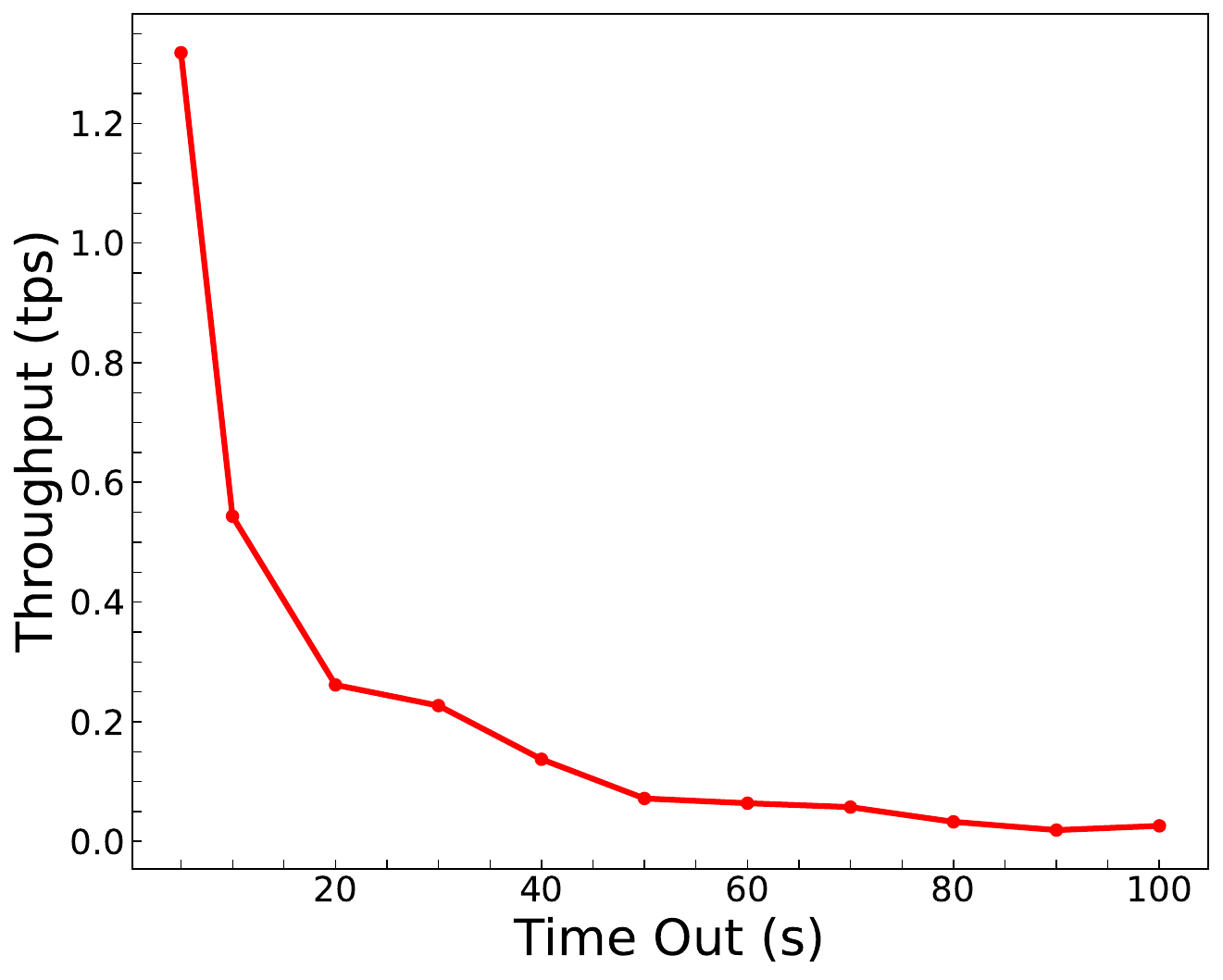}\label{fig:Tim_TP}} 
\vspace{-0.2cm}
\caption{Case Study 02 Results Varying Block Size and Timeout -
(a) Usage by Changing Block
(b) Usage by Changing Timeout
(c) Block size Activation
(d) Timeout Activation 
(e) Drop Rate - Changing the Block
(f) Drop Rate - Changing Timeout
(g) MRT - Changing Block
(h) MRT - Changing Timeout
(i) Flow - Changing Block
(j) Flow - Changing Timeout
}
\label{fig:sc2}
\end{figure*}

In the second case study, we try to observe the behavior of metrics from two perspectives.
First we vary the block size by setting a high \textit{timeout} (\texttt{TIME\_OUT} = 100s).
Second, we vary the \textit{block size} by setting a (\texttt{BLOCK} = 10).
The purpose of this case study is to individually analyze the impact of each of the two parameters.
Thus, we fixed the arrival rate at 100 tps and used other parameters following Table~\ref{tab:cenarioC}.
Figure \ref{fig:sc2} presents the results of this case study.

For utilization (See Figures \ref{fig:sc2}a and \ref{fig:sc2}b), it can be seen that in the endorsement and ordering steps, there is an increase in utilization proportional to the increase in the block size and \textit{timeout} value.
However, the \textit{timeout} variation leads to peak faster than block variation, which means that the timeout has a greater impact on utilization for such parameters than the block size.
The committer utilization drops due to the previous bottleneck because fewer transactions arrive at this stage.
Again, the drop in the \textit{committer} utilization by \textit{timeout} variation happens earlier than with block size variation because the variation of \textit{timeout} exhausts resources earlier.
In turn, the decrease in the \textit{committer} utilization in the block variation is more abrupt (from 100 to 0\%).
The trigger rate (See Figures \ref{fig:sc2}c and \ref{fig:sc2}d) focuses on one parameter at a time; the first parameter always has zero as the trigger rate, while the other varies as plotted.
When the bottleneck is reached, the trigger rate is reduced.
The trigger rate in block variation is zero, with a block size equal to 7.
In the timeout variation, the activation rate goes to zero with a timeout equal to 50s.
It is worth noting that a block size equal to 7 and a timeout equal to 50s are inflection points in all graphs, meaning that the respective value becomes null or peaks at these points.
The discard rate (PD) and the MRT increase as the bottleneck grows because the queuing waiting time increases.
PD reached 80\% in both cases (See Figures \ref{fig:sc2}e and \ref{fig:sc2}f).
The MRT (See Figures \ref{fig:sc2}g and \ref{fig:sc2}h) varied between 5s and ~60s.
The throughput (TP) (\ref{fig:sc2}i and \ref{fig:sc2}j), similar to the committer utilization, drops as the bottleneck grows.
The initial TP value is 70 tps.
This value is below the arrival rate (100 tps) because the system has some overhead in all its subcomponents.
To verify this, observe the use of the commit at 100\% in the initial points (See Figures \ref{fig:sc2}a and \ref{fig:sc2}b).

For this second case study, we may highlight that both parameters (block size and timeout) greatly impact the system behavior.
The increase in such parameters causes significant bottlenecks in the ordering step.
This bottleneck increases until it reaches a maximum point that we call the inflection point, which for this study was caused by a block size equal to 7 and a \textit{timeout} greater or equal to 50s.
Changing the block size by just one unit can greatly change the MRT.
Changing the block size from 6 to 7 changed the MRT from 5 to 60 seconds.

\subsection{Case Study 03 - Interaction between Block and TimeOut}

In the previous case study, we observed the behavior of the two parameters evaluated block size and timeout independently.
This way, it was possible to perceive how much each parameter directly interfered with the performance metrics.
%

When we previously varied the block size, we fixed \textit{timeout} at 10s, so the trigger rate per \textit{timeout} remained null.

\begin{figure}[!htb]
    \centering
    \includegraphics[width=3in]{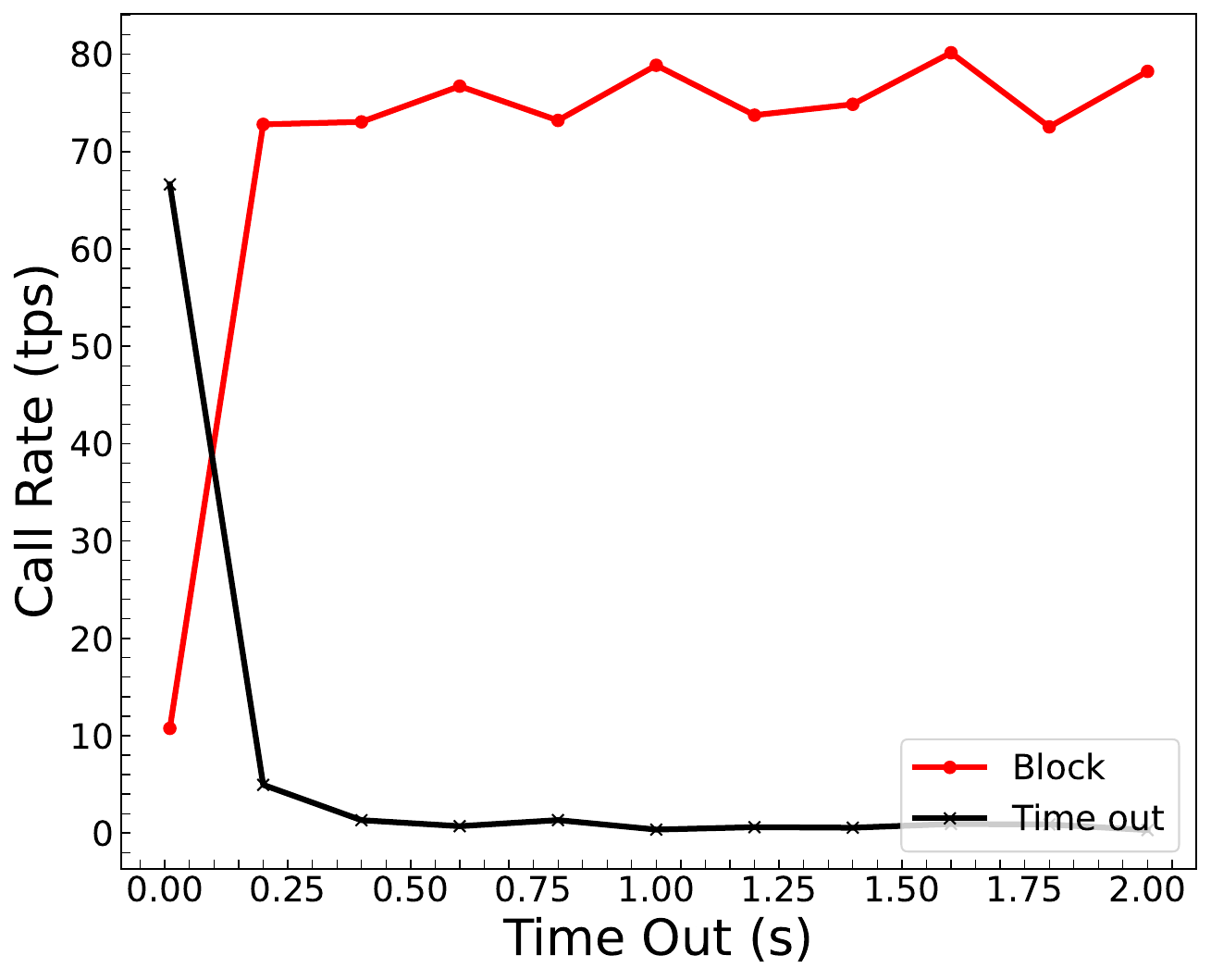}
\caption{Interaction between Block and Timeout}
\label{fig:Toff}
\end{figure}

Now, in the third case study, we vary the \textit{timeout} from zero to 2s, fixing the block size with a lower value (\texttt{BLOCK} = 6).
The objective this time is to obtain an interaction result between the two activation rates.
Figure \ref{fig:Toff} presents such a result.

The graph shows an intersection between the two lines.
As the timeout increases, the activation rate per timeout decreases as this path becomes more restrictive, meaning that it will take longer to trigger by timeout. Timeout taking longer gives more chances for a complete block to be generated.
Blockchain network researchers report that it is somewhat complex to calibrate these two parameters precisely because of this possible \cite{thakkar2018performance,sukhwani2018performance} interaction.
The present model contributes, therefore, to this prediction of system behavior.
For this third case study, some interesting conclusions can be made, such as:
\begin{enumerate}
    \item The model allows determining the point where the two lines intersect each other, that is, the point where more complete blocks are formed than partial blocks;
    \item the trigger rate per block reaches stability at a certain point because the system has reached high queuing;
    \item for a \texttt{TIME\_OUT} = 0, mostly partial blocks will be formed;
    \item for a \texttt{TIME\_OUT} = 2s, mostly complete blocks (of size 6) will be formed.
\end{enumerate}

The changes between complete and incomplete block size can increase the variability of the transaction time, making it difficult to predict. It is up to permissioned blockchain network administrators to configure the block size and \textit{timeout} based on the transaction rate observed on the network.

\subsection{Case Study 04 - Sensitivity Analysis}

As an extension of the previous case study, in this one, we try to quantify the impact of each component separately and their correspondent interaction over the system performance; based on a $2^k$ DoE, we established a Sensitivity Analysis that enable us to identify which component impacts the most on the MRT.

\begin{figure}[!htb]
    \centering
        \includegraphics[width=8cm]{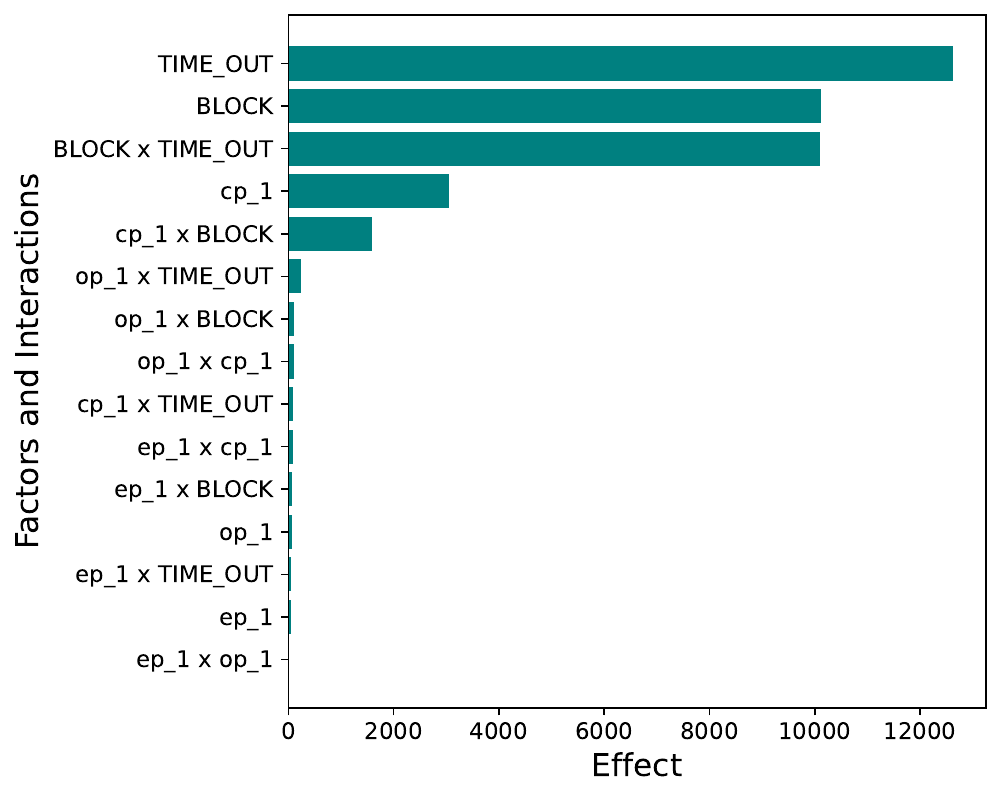}
     \caption{Factors and Interactions and their respective effect.}
  \label{fig:mod}
\end{figure}

Figure \ref{fig:mod} presents the overall effect plot for each factor and their respective interaction and effect on the metric of interest. As can be seen in this plot, the \texttt{TIME\_OUT} has the biggest impact over the metric of interest, while \texttt{block size} and their respective interaction with the \texttt{TIME\_OUT} follows close by.

\begin{figure}
    \centering
    
    \begin{minipage}{1\textwidth}
        \centering
        \begin{subfigure}[b]{0.25\textwidth}
  \includegraphics[width=\linewidth]{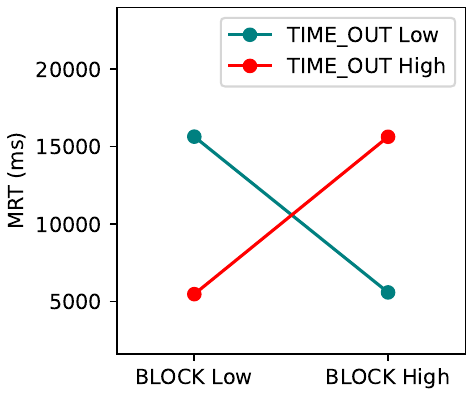}
  \label{fig:a}\caption{}
        \end{subfigure}
        \begin{subfigure}[b]{0.25\textwidth}
  \includegraphics[width=\linewidth]{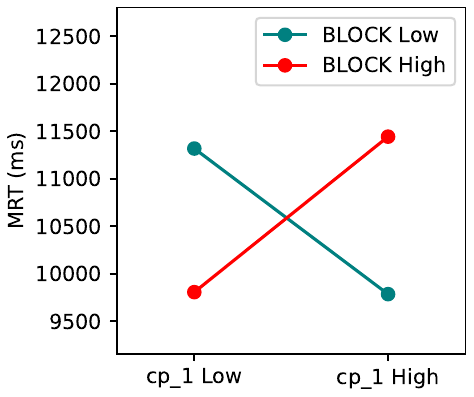}
  \label{fig:b}\caption{}
        \end{subfigure}
        \begin{subfigure}[b]{0.25\textwidth}
  \includegraphics[width=\linewidth]{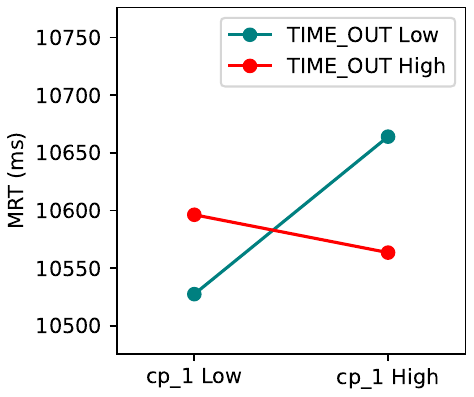}
  \label{fig:c}\caption{}
        \end{subfigure}
        \begin{subfigure}[b]{0.25\textwidth}
  \includegraphics[width=\linewidth]{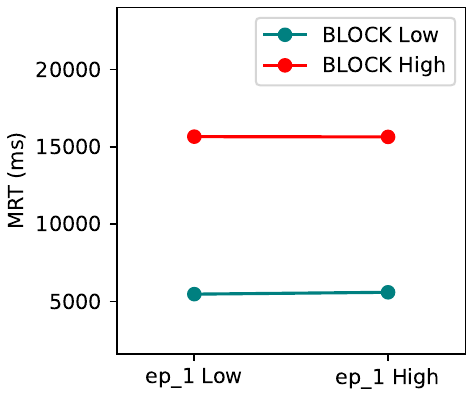}
  \label{fig:d}\caption{}
        \end{subfigure}
        \begin{subfigure}[b]{0.25\textwidth}
  \includegraphics[width=\linewidth]{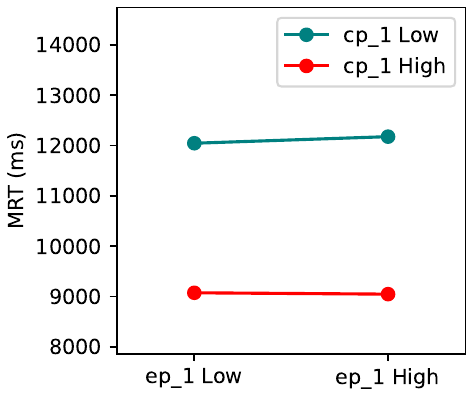}
  \label{fig:e}\caption{}
        \end{subfigure}
        \begin{subfigure}[b]{0.25\textwidth}
  \includegraphics[width=\linewidth]{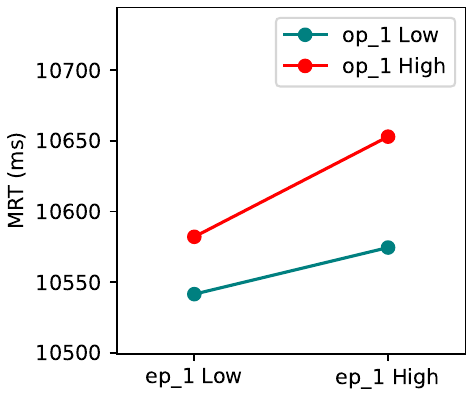}
  \label{fig:f}\caption{}
        \end{subfigure}
        \begin{subfigure}[b]{0.25\textwidth}
  \includegraphics[width=\linewidth]{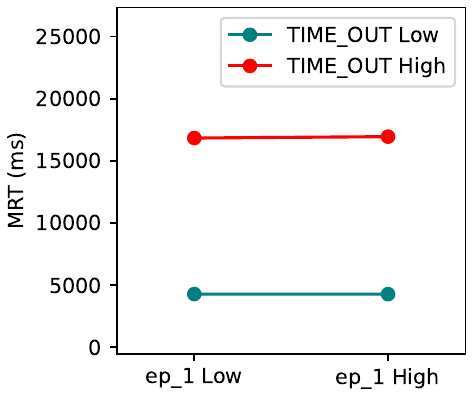}
  \label{fig:g}\caption{}
        \end{subfigure}
        \begin{subfigure}[b]{0.25\textwidth}
  \includegraphics[width=\linewidth]{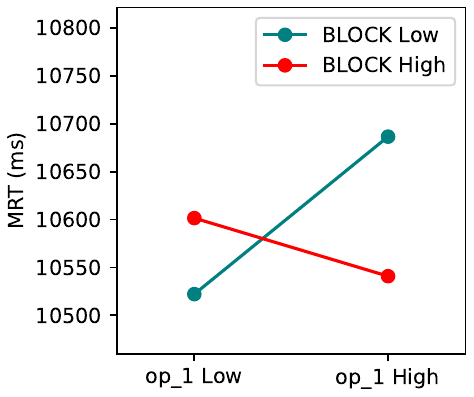}
  \label{fig:h}\caption{}
        \end{subfigure}
        \begin{subfigure}[b]{0.25\textwidth}
  \includegraphics[width=\linewidth]{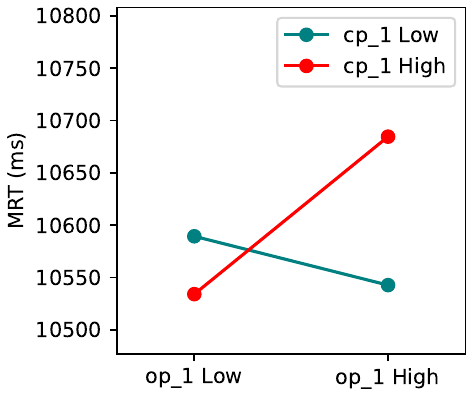}
  \label{fig:i}\caption{}
        \end{subfigure}
        \begin{subfigure}[b]{0.25\textwidth}
  \includegraphics[width=\linewidth]{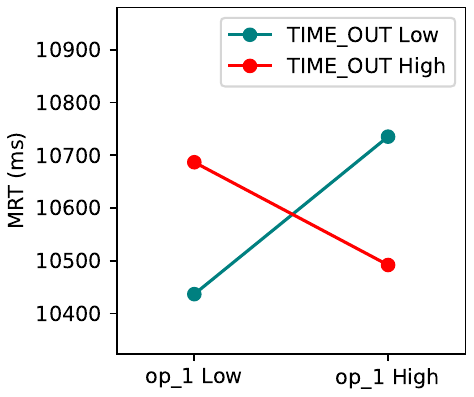}
  \label{fig:j}\caption{}
        \end{subfigure}
        \label{fig:linha2}
    \end{minipage}
\vspace{-0.2cm}
\caption{Interaction Plot -
(a) Block size vs. timeout
(b) Block size vs. cp\_1
(c) Timeout vs. cp\_1
(d) ep\_1 vs. block size
(e) ep\_1 vs. cp\_1
(f) ep\_1 vs. op\_1
(g) ep\_1 vs. timeout
(h) op\_1 vs. block size
(i) op\_1 vs. cp\_1
(j) op\_1 vs. timeout
}
    \label{fig:figura_com_subfiguras}
\end{figure}

The values varied from \textbf{low} to \textbf{high}, indicating a minimum and a maximum value for each factor ($2^k$). Figure \ref{fig:figura_com_subfiguras} presents the strength of each relationship. Some factors are unrelated, meaning their lines walk parallel, never touching each other.

As presented by plots (See Figures \ref{fig:figura_com_subfiguras}d, e, f, g), the unrelated factors present a set of relationships that can receive less attention from system administrators. At the same time, some other relationships can be better explored to improve system performance regarding MRT.

As for the main conclusions of this last case study, we may again assume that the relationship between timeout and block size requires some depth evaluation (See Figure \ref{fig:figura_com_subfiguras}a). From this point of view, the MRT can move from 5s to more than 15s, a more than 200\% increase. This impact is overwhelming not for the system but for its users. Expecting a credit card transaction to take 15 seconds nowadays is too much; the same applies to any blockchain transaction.



\section{Conclusions and Future Works}
\label{sec:conclusão}

In this work, we proposed an SPN model to analyze the performance of a permissioned blockchain on the Hyperledger Fabric platform. 
The model evaluates critical performance indicators such as mean response time, throughput, utilization, and transaction discard probability, providing practical insights into the system's efficiency under various blockchain parameters like block size, timeout, and transaction arrival rate. 
We explored different parameter variations related to resource capacity, queuing, and processing resources to pinpoint areas for improving system performance.
Nonetheless, the four case studies highlight trade-offs between block configurations (block size and timeout) and the computational capacity of architectural components affecting network delay and throughput. 
These case studies provide practical insights for analyzing Hyperledger Fabric platform performance.
However, it's important to note the limitations of our study: (i) we did not conduct experimental evaluations to validate the model results; (ii) the evaluated scenarios do not consider the use of different databases, such as CouchDB or LevelDB, that may significantly impact the overall performance; (iii) we do not consider the gossip protocol in the modeled environment, which could extend the state space due to the large number of messages exchanged between all nodes and compromise the model evaluation.
Future iterations of the model will incorporate decision weighting, explore the use of different databases, validate the model's results through experimental evaluations, and prioritize protocol steps among machines in the blockchain network, opening up new avenues for research and practical application. 
To enhance our understanding further, we aim to explore this functionality in a case study.


\bibliography{sn-bibliography}

\end{document}